\documentclass{aastex63}

\received{June 1, 2019}
\revised{January 10, 2019}
\accepted{\today}
\submitjournal{AJ}

\shorttitle{WD in PN}
\shortauthors{Ahumada et al.}

\begin{document}

\title{Hunting young white dwarfs at the center of
planetary nebulae}

\correspondingauthor{Javier A.\ Ahumada}
\email{javier.ahumada@unc.edu.ar}

\author{Javier A.\ Ahumada}
\affiliation{Observatorio Astron\'{o}mico, Universidad Nacional de C\'{o}rdoba, Laprida 854, 5000 C\'{o}rdoba, Argentina}
 
\author{Walter A.\ Weidmann}
\affiliation{Observatorio Astron\'{o}mico, Universidad Nacional de C\'{o}rdoba, Laprida 854, 5000 C\'{o}rdoba, Argentina}
\affiliation{Consejo Nacional de Investigaciones Cient\'{\i}ficas y T\'ecnicas, Argentina}

\author{Marcelo M. Miller Bertolami}
\affiliation{Facultad de Ciencias Astron\'omicas y Geof\'{\i}sicas,
           Universidad Nacional de La Plata,\\
           Paseo del  Bosque s/n,
           (1900) La Plata, Argentina}
\affiliation{ Instituto de Astrof\'{\i}sica La Plata, IALP, CONICET-UNLP, \\Paseo del  Bosque s/n, (1900) La Plata, Argentina }       
\author{Leila Saker}
\affiliation{Observatorio Astron\'{o}mico, Universidad Nacional de C\'{o}rdoba, Laprida 854, 5000 C\'{o}rdoba, Argentina}




\begin{abstract}

We present Gemini-South observations
 of  nine faint and extended planetary nebulae.
 Using direct images taken with the spectrograph GMOS,
we  built the $(u' - g')$ vs.\ $(g' - r')$ diagrams of the stars in the observed areas 
which allowed us, also considering their geometrical positions,  
to identify the probable central stars  of the nebulae.
 Our  stellar spectra of seven stars, also
taken with GMOS, indicate that four (and probably two more) objects
are white dwarfs of the 
DAO subtype. Moreover, the white dwarf status of the four stars is confirmed
by   the  parameters $ T_{\mathrm{eff}}$ and  $ \log g$
derived with the help of theoretical stellar spectra. 
Given this evidence, 
we propose that these hot stars are  the central ionizing sources of the nebulae.
With this work  we hope to help improve  the current 
 scarce statistics on central white dwarfs in
planetary nebulae.

\end{abstract}





\section{Young white dwarfs: The ionizing sources of old planetary nebulae}
\label{sec:intro}

Planetary nebulae (PNe) are old, evolved objects,  in which 
 the slowly-expanding, surrounding gas that  an 
intermediate- or low-mass star  has
 lost during the asymptotic giant branch (AGB) phase, is
 ionized by the ultraviolet flux emanating from the central star.
The nebula emits mainly 
in the ultraviolet, visible, and IR spectral regions.
The central star of the nebula (CSPN) is  essentially the 
hot stellar  core plus what is left of its
  initial envelope (\citealt{na98}, \citealt{dr99}). While the 
morphologies of the PNe are amazingly diverse 
\citep{ww16}, the CSPNe  also have a wide 
range of properties, with temperatures
that range from \textbf{$\mathbf{\approx25}$,000} to over 200,000~K,  luminosities
 from 10 to over 10,000~$L_\odot$, and a display of astonishingly 
varied  spectra.
Although most  CSPNe
 have hydrogen-rich atmospheres \citep{to06}, there is, however,
another class of central stars with atmospheres deficient in hydrogen.
Some H-poor CSPNe have strong emission lines of highly ionized carbon, 
oxygen, and helium, and exhibit fast stellar winds---that produce broad emission 
lines---as well as high mass-loss rates \citep{we08}. Finally, there is a small group of 
CSPNe   showing rare spectral types like, 
e.g., O(He). As yet
 the evolution of the CSPNe is not fully understood, in particular that of 
the H-deficient ones. It is generally accepted that this evolutionary path begins with a late 
O-type star or a [WCL] star, depending on whether it is  H-rich or  H-poor, and 
   ends up with a
white dwarf  (WD), of DA-type, or of DO-/DB-type, 
respectively \citep{na98}.  A subclass of the DAs is that of the DAOs, 
 whose spectra  are characterized by broad
Balmer lines together with a sharp \ion{He}{2} feature at~4686~\AA\ \citep{wes85}.
In  the H-poor sequence, before the star   
becomes a WD,  it is believed it goes through the PG~1159 stage. This is
represented by  stars showing a strong  \ion{He}{2}/\ion{C}{4}
absorption trough around 4670~\AA\ and typical temperatures of 100,000~K. The
 PG~1159 stars are key objects for a fully understanding of the post-AGB evolution \citep{dr99}.
At present, although there are 3000 true and probable PNe known in the 
Milky Way,  a stellar continuum has been detected in only 16\% of them 
\citep{we11}. Less than 6\% of these CSPNe are classified as WDs 
(i.e., 30~stars, of which 12 are DA-type, 14 are DAO, and 4 are DO) 
and, in most cases, 
they have been identified and classified  from low-resolution  spectra.
Their study is increasingly difficult after the PN reaches its
maximum brightness (i.e., approximately when nuclear fusion stops in the CSPN), 
because   the central star grows fainter while evolving down the white dwarf cooling 
track. The star  finally turns undetectable  like the PN,
that by this time has become very dispersed.

This work is part of a current effort to unveil the properties of
faint, so far unstudied CSPNe to gain  knowledge on these
stars and their evolution \citep{we11}.
In particular we are interested in finding new WDs 
among CSPNe, because so few of them are currently known. 
Since the WDs are intrinsically faint objects, large telescopes
are required for their detection and proper study. Here we use photometric 
and spectroscopic data acquired with the 8.1-meter Gemini-South Telescope,
together with theoretical spectra and evolution models,
to  find and characterize
the central white dwarfs of seven planetary nebulae: 
\object[PN G019.7$-$10.7]{PN~G019.7$-$10.7},
\object[PN G237.0$+$00.7]{PN~G237.0$+$00.7},
\object[PN G276.2$-$06.6]{PN~G276.2$-$06.6},
\object[PN G298.7$-$07.5]{PN~G298.7$-$07.5},
\object[PN G302.1$+$00.3]{PN~G302.1$+$00.3},
\object[PN G314.5$-$01.0]{PN~G314.5$-$01.0}, and
\object[PN G325.3$-$02.9]{PN~G325.3$-$02.9}.
We also present the probable central stars of
 two additional nebulae,
\object[PN G328.5$+$06.0]{PN~G328.5$+$06.0} and
\object[PN G344.9$+$03.0]{PN~G344.9$+$03.0},
for which we only possess Gemini photometry.

The plan of the paper is as follows:  
In Section~\ref{sec:targets} we justify our choice of the 
planetary nebulae and describe how  we intend to find their  central stars.
In  Section~\ref{sec:obs} we give an account of 
the  photometric and spectroscopic observations 
performed with GMOS at the Gemini-South Telescope.
In Section~\ref{sec:fot} we explain the photometric analysis and the identification of the  CSPNe.
In Section~\ref{sec:esp} we present and discuss our spectroscopy of seven stars, obtain
 parameters $T_\mathrm{eff}$ and $\log g$ for four of them employing TheoSSA synthetic spectra,
 and briefly examine some properties of hot subdwarfs stars as CSPNe. 
 Colors and distances for the CSPNe derived from the \emph{Gaia} 
 DR2 survey are  discussed in Section~\ref{sec:gaia}.
 Using stellar evolution model sequences, in Section~\ref{sec:evolucion}, 
 physical parameters  of four CSPNe are computed.
Summary and  conclusions are in  Section~\ref{sec:sum}. 
In  Appendix~\ref{app:apendice} we present
finding charts for the nine stars.


\section{Targets selection}
\label{sec:targets}

As suitable objects for searching new WDs as CSPNe,
we selected nine planetary nebulae from the MASH 
catalogs\footnote{\url{https://heasarc.gsfc.nasa.gov/W3Browse/all/mashpncat.html}}
 (\citealt{maI06}, \citealt{maII08}). The PNe of our sample, listed in
Table~\ref{tab:np}, have  large angular sizes,  symmetrical
 morphologies, and low surface brightness. 
We expect that objects with these properties are
indeed old PNe, and therefore it is highly probable that their CSPNe are  
evolved objects, like white dwarfs.
The MASH catalogs give the following comments on  these nebulae:

    
\noindent \emph{PN~G237.0$+$00.7:} Very large, very faint, vaguely circular nebula; 
has strong [\ion{O}{3}] and no H$\beta$, 
[\ion{N}{2}]~$\approx$~H$\alpha$; probable CSPN position used for nebula centre.

\noindent \emph{PN~G276.2$-$06.6:}     Lovely faint, large, circular, evolved shell PN 
with enhanced NW edge; has [\ion{N}{2}]~$\approx 2\times$H$\alpha$, [\ion{O}{3}]~$>$~H$\beta$;
has probable CSPN.

\noindent \emph{PN~G298.7$-$07.5:} Very large, fractured oval nebula, possible evolved PN; 
also observed SA290603, MS060105; [\ion{N}{2}]~$>10 \times$H$\alpha$, 
[\ion{O}{3}]~$>$~H$\beta$; CSPN at 12:02:55.5, $-$70:00:57.

\noindent \emph{PN~G302.1$+$00.3:} Bright, large, bipolar like structure, previously known 
as \textbf{\ion{H}{2}} region RCW69; confirmed PN, has [\ion{N}{2}]~$>$~H$\alpha$; also 
observed SA240603; was PHR1244$-$6230.

\noindent \emph{PN~G314.5$-$01.0:} Faint, extensive PN; confirmed by spectra.
[\ion{N}{2}]~$\gtrsim 2.5\times$H$\alpha$, [\ion{O}{3}]~$>$~H$\beta$, CSPN at 14:32:09.7, $-$61:38:41; 
previously NUN NeVe GN~14.28.3.01; possible IRAS source 14281$-$6127.

\noindent \emph{PN~G325.3$-$02.9:} Area of diffuse emission; [\ion{O}{3}]~$\approx$~$5\times$H$\beta$, 
\textbf{\ion{He}{2}}, H$\alpha$ only in red.

\noindent \emph{PN~G328.5$+$06.0:} Very faint ring nebula, [\ion{N}{2}]~$\approx$~$4\times$H$\alpha$,
[\ion{O}{3}]~$>$~H$\beta$.

\noindent \emph{PN~G344.9+03.0:} PHR1626$-$5216 analogue with striations.
  
Four probable CSPNe are reported by the MASH catalogs, found
 using their  blue ($B_J - R_F<0$) colors. We will use these
identifications to check the results of our own strategy, which we describe next.

To identify the CSPNe among the stars present in the observed fields 
we  use two criteria, namely, photometric and geometric.
First, by performing photometry on the sky areas centered on the nebulae (Section~\ref{sec:fot}), 
we build the color-color $(u'-g')$ vs.\ $(g'-r')$ diagram. 
As shown by \cite{gir11},
 in this plane the white dwarfs lie along a sequence clearly differentiated
from that of main-sequence stars. In Figure~2 \textbf{(left panel)} of \citeauthor{gir11},
the white-dwarf area is 
delimited with a line, and the knee of the upper main sequence appears at the 
bottom of the \textbf{plot; for comparison,} there is  a calibrated main sequence 
  in Figure~4 of \cite{bil08}.
 We then expect to find WD candidates, if there are any, in this part
 of the diagrams. An objection to the use of our color-color diagrams could be raised, namely, that
they are built with instrumental
magnitudes; this is so because  Band~4 nights, under which the photometric observations 
 were performed (Section~\ref{sec:imaging}),
 are by definition non-photometric. This was done on purpose,  since we are interested only 
 in the general shapes of the sequences, and
 we expect that they
 do not change significantly in the instrumental system. Even the
presence of thin clouds should only change the photometric zero points
and not the shapes of the sequences.
  
The second criterion of selection,  the geometrical one, complements the photometric: once a
 WD candidate is found in the color-color diagram, we  check its
position in the image. The star we are looking for should lie, in principle, at the geometrical
center of the nebula---although there are known exceptions like, for example, the central star 
of \object[Sh 2-174]{Sh~2-174}, see Figure~5
of \cite{na98}. A nebula with symmetric, rounded morphology  makes easier the location of its center,
 helping us give more weight to this criterion.

\begin{deluxetable*}{lccccccc} 
\tablecaption{Data of the planetary nebulae and of their proposed central stars
\label{tab:np}}
\tablewidth{0pt}
\tablehead{%
\multicolumn{5}{c}{PN} & \multicolumn{3}{c}{CSPN} \\\hline
\colhead{Usual name} &
\colhead{PN~G} &
\colhead{Dimension} &
\colhead{Morph\tablenotemark{a}} & 
\colhead{$\log F_{\mathrm{red}}$\tablenotemark{b}} &
\colhead{$\alpha$} &
\colhead{$\delta$} &
\colhead{$r'$\tablenotemark{c}} \\
 &&   (arcsec) & & (mW~m$^{-2}$) & (h~m~s)  & (\arcdeg~\arcmin~\arcsec) & (mag)
}
\startdata
MPA~J1906$-$1634 & 019.7$-$10.7 &  242 & B   & $-$11.57 & 
19 06 32.79 &  $-16$ 34 00.2     & 18.8 \\ 
PHR~J0740$-$2055 & 237.0$+$00.7  &   240 & Ra  & $-$12.19 & 07 40 22.91 
&  $-20$ 55 54.5 &  18.1\\ 
PHR~J0907$-$5722 & 276.2$-$06.6  &    241 & Rsm & \nodata & 
09 07 51.02  &$-57$ 22 53.3 & 18.7 \\ 
PHR~J1202$-$7000 & 298.7$-$07.5 &   317 & Eas & $-$11.20 & 
12 02 55.46  & $-70$ 00 56.8 & 19.7 \\ 
RCW~69           & 302.1$+$00.3 &   300 & B   & $-$10.40 & 
12 44 27.53  &$-62$ 31 19.2 &  19.1 \\ 
PHR~J1432$-$6138 & 314.5$-$01.0 &   180 & Es  & $-$10.90 & 
14 32 09.64     & $-61$ 38 41.4       &  18.6 \\ 
PHR~J1553$-$5738 & 325.3$-$02.9 &   133 & E   & $-$11.45 & 
15 53 09.85  &$-57$ 38 06.0 & 16.5 \\  
 PHR~J1533$-$4834 & 328.5$+$06.0 &   162 & Rr  & \nodata  &    
 15 33 34.06   & $-48$ 34 24.1      &  18.6\\ 
BMP~J1651$-$3930 & 344.9+03.0  &  315 & Eas/Isa & \nodata & 
16 51 41.27  & $-39$ 30 27.5 & 19.0 \\ 
\enddata
\tablenotetext{a}{Morphological classification, from the MASH catalogs (\citealt{maI06}, \citealt{maII08}).}
\tablenotetext{b}{Average red flux (H$\alpha$+[\ion{N}{2}]),  
\cite{2013MNRAS.431....2F}.} 
\tablenotetext{c}{Estimated value.}
\end{deluxetable*}


\section{Observations and data reduction}
 \label{sec:obs}

 Our observations comprise, first,  broad-band direct imaging of the 
 nine nebulae, aimed
at the identification of their ionizing stars; second,
 long-slit spectroscopy of such stars
  for their
spectroscopic classification.  
 
\subsection{Broad-band optical imaging}
\label{sec:imaging}

Direct images of  the PNe of Table~\ref{tab:np}
were taken  with the
Gemini-South multi-object spectrograph\footnote{\url{http://www.gemini.edu/sciops/instruments/gmos-0}}
 (GMOS) in its imaging mode, mounted on the 8.1-meter Gemini-South Telescope at Cerro Pach\'{o}n, Chile. 
The detector is an array
of three $2048\times4176$ Hamamatsu chips arranged in a row.
  The array was configured in a $2\times2$ binning mode, which gives a scale of 0.16\arcsec/pixel. 
  The useful field
  covers a sky area of $5.5\times 5.5$~arcmin$^2$.
  The programs under which these data were obtained are GS-2016B-Q-92,
 GS-2017A-Q-90, and GS-2018A-Q-404
 (PI:~Weidmann), and were intended for Band~4 or ``poor weather'' nights, since the expected quality
 was considered sufficient for this part of the work (cf.~Section~\ref{sec:targets}).

The filters utilized were broad-band $u'$, $g'$, and $r'$, 
similar to those of the Sloan Digital Sky Survey \citep{fuk96}.
 Per filter, several images were taken and combined after their processing  with the Gemini IRAF 
package.\footnote{\url{http://www.gemini.edu/sciops/data-and-results/processing-software}}
The number of exposures and the integration times were, in all cases, the following:
 $4\times675$~s ($u'$), $8\times67$~s ($g'$),
 and $20\times60$~s ($r'$).  
 With these exposure times it should be possible to observe  
an O-type star of magnitude $V=19.5$ with a signal-to-noise (S/N) ratio
of about~5. 
 The quality of our final combined frames can be assessed by the mean 
FWHM of the stars in each band, which are shown in Table~\ref{tab:datos_observacionales}.

\begin{deluxetable*}{lcccc} 
\tablecaption{Some remarks on the observations
\label{tab:datos_observacionales}}
\tablewidth{0pt}
\tablehead{
\colhead{PN~G} &
\multicolumn{3}{c}{Mean FWHM\tablenotemark{a}}  &
Exposures\tablenotemark{b} \\ \cline{2-4}
 & $u'$ & $g'$ & $r'$ & 
}
\startdata
019.7$-$10.7 & 1.06   & 0.80   & 1.04 & $3\times760$\\
237.0$+$00.7 & $1.90$  &  1.84 & 1.79 & $3\times300$  \\  
276.2$-$06.6 & 1.98  & 1.97  & 1.79 & $3\times450$ \\  
298.7$-$07.5 & 1.04  & 1.20  & 1.00 & $3\times1200$ \\
302.1$+$00.3 & 1.52  & 1.58  & 1.36 & $3\times925$  \\
314.5$-$01.0 & 1.70 & 1.12 & 1.20 & $3\times650$  \\ 
325.3$-$02.9 & 1.38  & 0.88  & 1.50 & $3\times1100$ \\
328.5$+$06.0 & 1.47 & 1.30 & 1.28 & \nodata \\  
344.9$+$03.0 & 1.14 & 1.09& 1.07& \nodata \\
\enddata
\tablenotetext{a}{For direct imaging, in arcseconds.}
\tablenotetext{b}{For spectroscopy data, in seconds.}
\end{deluxetable*}

 
\subsection{Spectroscopy data}
\label{essp} 

The data for spectroscopy were gathered under the Gemini programs 
GS-2017B-Q-80, 
GS-2018A-Q-301,
GS-2019A-Q-304, and
GS-2019A-Q-405
(PI: Weidmann).
Seven stars, out of the nine
 found as described in Section~\ref{sec:fot}, were observed with 
GMOS in its long-slit spectroscopic mode.
Diffraction grating B600 was used 
with a slit width of 1.0~arcsec, rendering a spectral resolution $\mathbf{R\approx 1400}$ and a
covered wavelength range that goes from 4000 to 7000~\AA.
The integration times for spectroscopy on source are given in
Table~\ref{tab:datos_observacionales}.

 Using IRAF\footnote{IRAF:
 the Image Reduction and Analysis Facility is distributed by the National Optical  
Astronomy Observatories, 
 which are operated by the Association of Universities for Research 
in Astronomy, Inc., under cooperative agreement with the National  Science Foundation.} tasks, 
the spectra were reduced following standard procedures:
 overscan and combined bias subtraction,
dome flat-fields, and cosmic ray removal. 
The technique employed to remove the nebular contribution is described in
 \citet{2018A&A...614A.135W}.
 Finally, the three spectra per target were averaged.


\section{Photometry}
\label{sec:fot}
  
 The photometry  was carried out with \textsc{daophot},  in IRAF.  
Simple aperture photometry was performed on the $u'$, $g'$, and $r'$ images, 
adopting small apertures of  1~FWHM of radius to achieve a good S/N relation; 
neither aperture correction nor  standard stars were required.
Fig.~\ref{fig:c-cd1}
shows the nine instrumental  $(u'-g')$ vs.\ $(g'-r')$ 
diagrams. In all of them the sequence of dwarfs is clearly seen and, in
the  zone of the white dwarfs (cf.~Section~\ref{sec:targets}), stars,
marked by arrows, stand out. Since the locations of these stars
indeed coincide with the centers of their respective nebulae,
 we propose that they are their   
ionizing sources. The J2000.0 coordinates of the nine stars are listed in Table~\ref{tab:np}
and the finding charts are shown in the Appendix~\ref{app:apendice}
(Figs.~\ref{fig:cartas} and~\ref{fig:cartas2}).

Our positions are equal  to the
 arcsecond for those CSPNe already indicated in the MASH
catalogs. This confirms our results, and gives us confidence in
the procedure we followed to do the detections. 
The  coordinates of the remaining stars, moreover, are also   very near to those
of the  PNe in the MASH catalogs; 
 this is a further corroboration that  our criterion of choosing  symmetrical nebulae was
essentially correct.

\begin{figure} 
\centering
\includegraphics[scale=0.34]{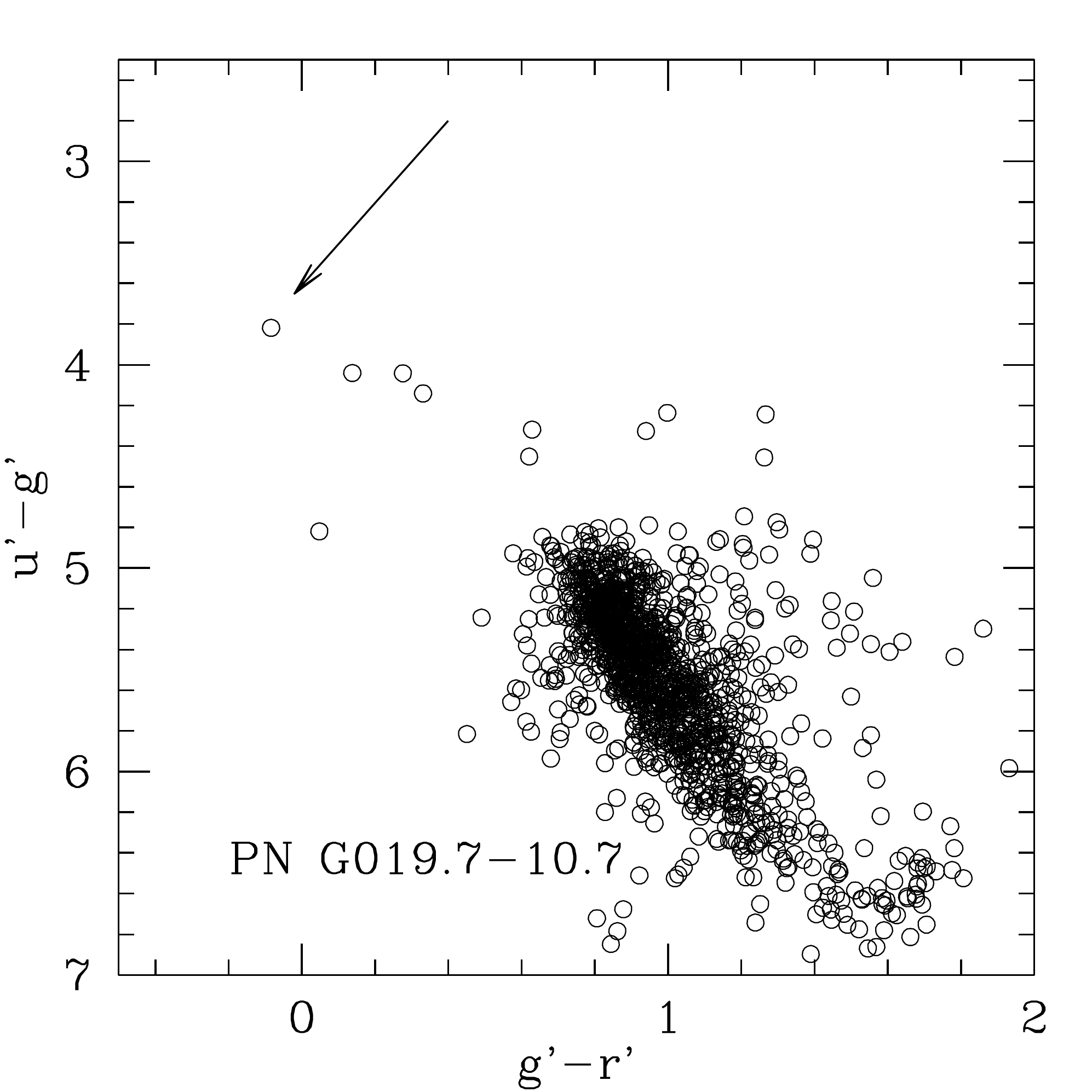}
\includegraphics[scale=0.34]{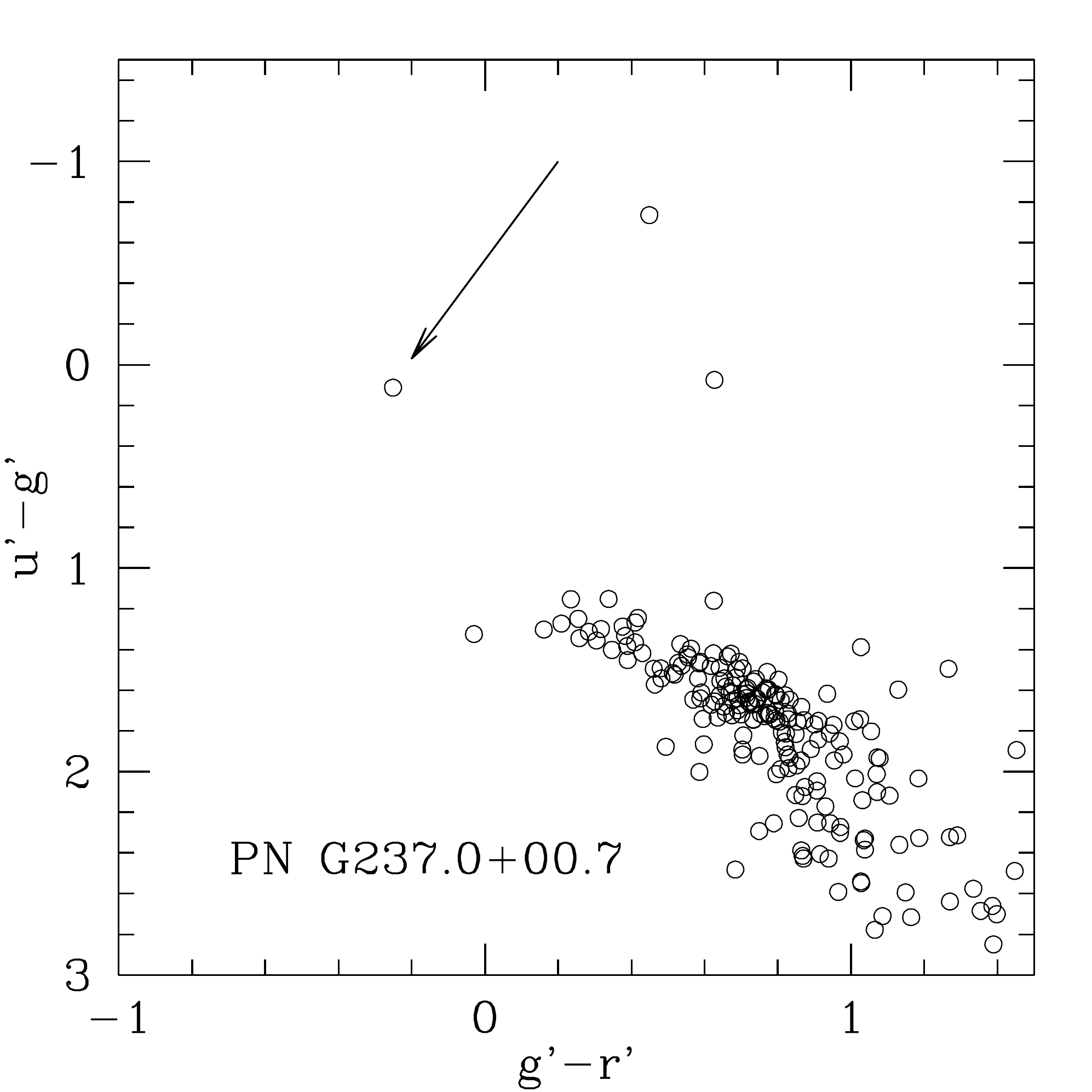}

\includegraphics[scale=0.34]{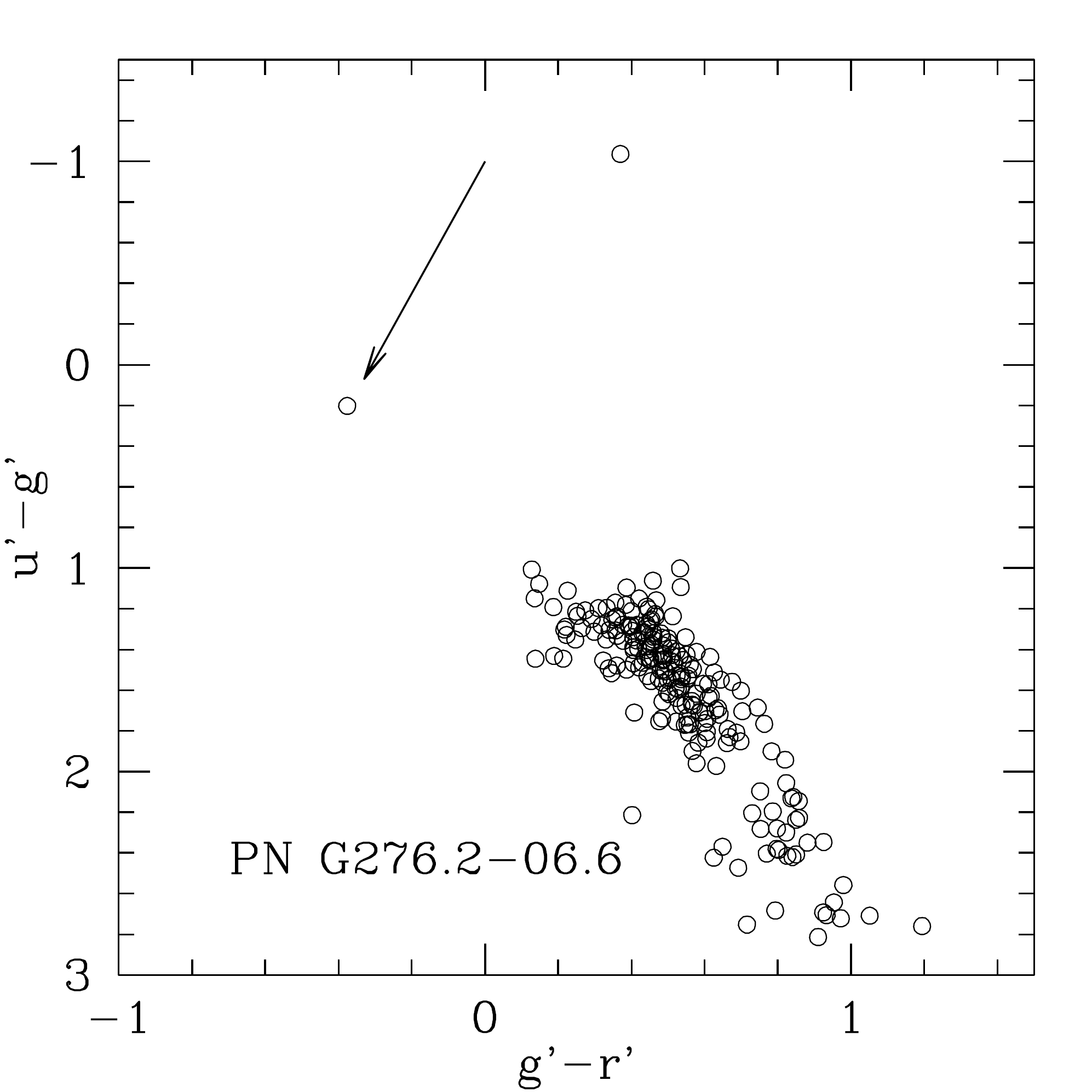}
\includegraphics[scale=0.34]{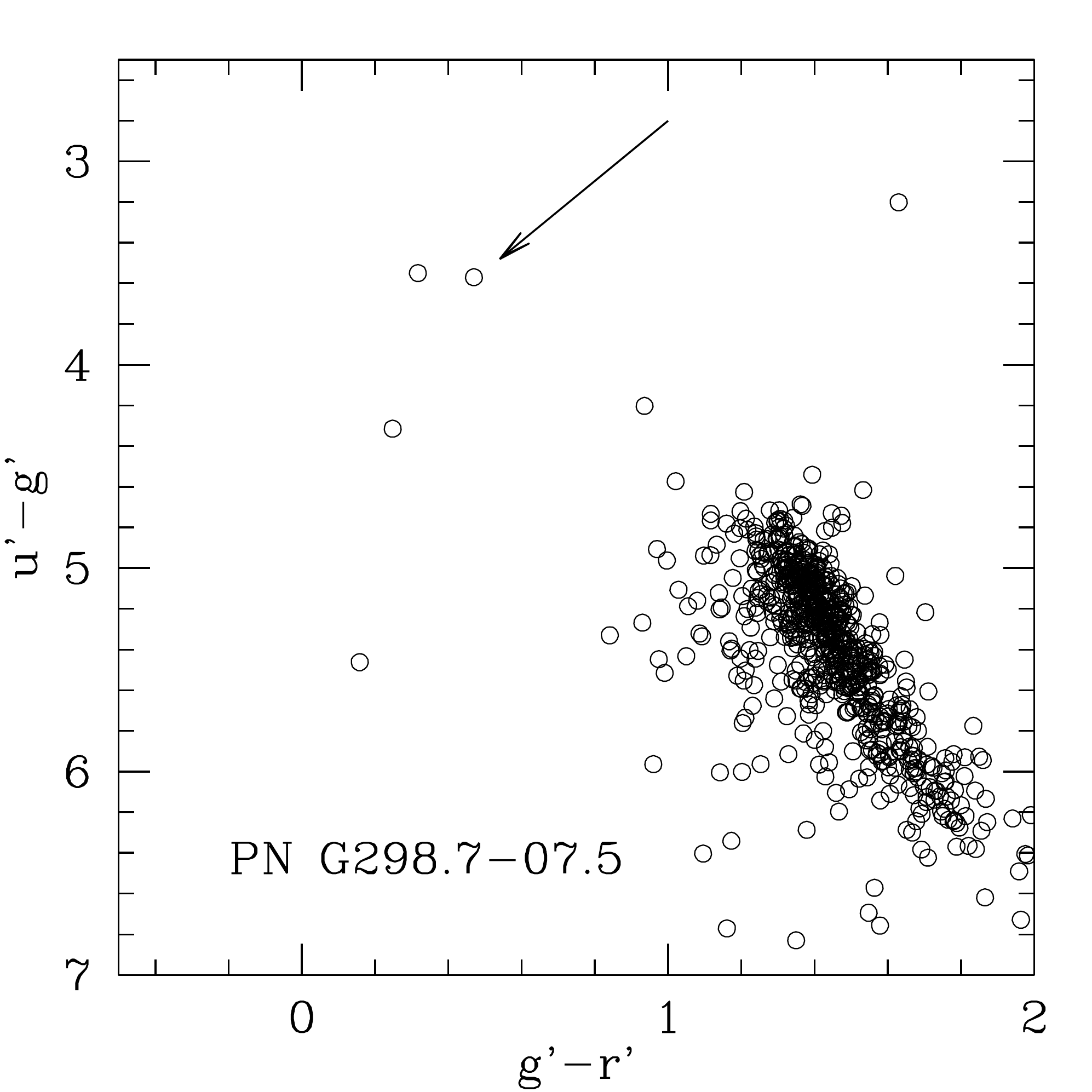}

\includegraphics[scale=0.34]{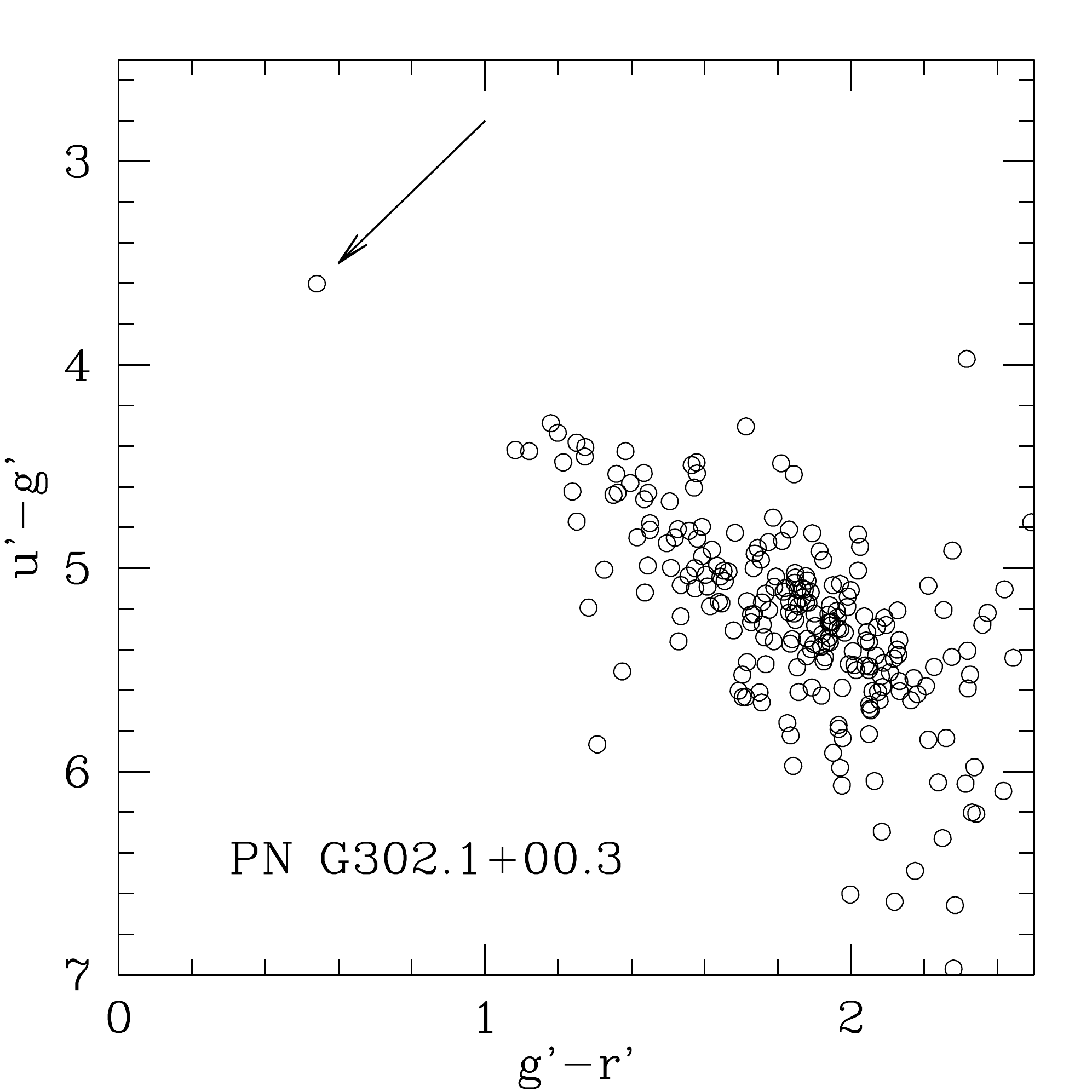}
\includegraphics[scale=0.34]{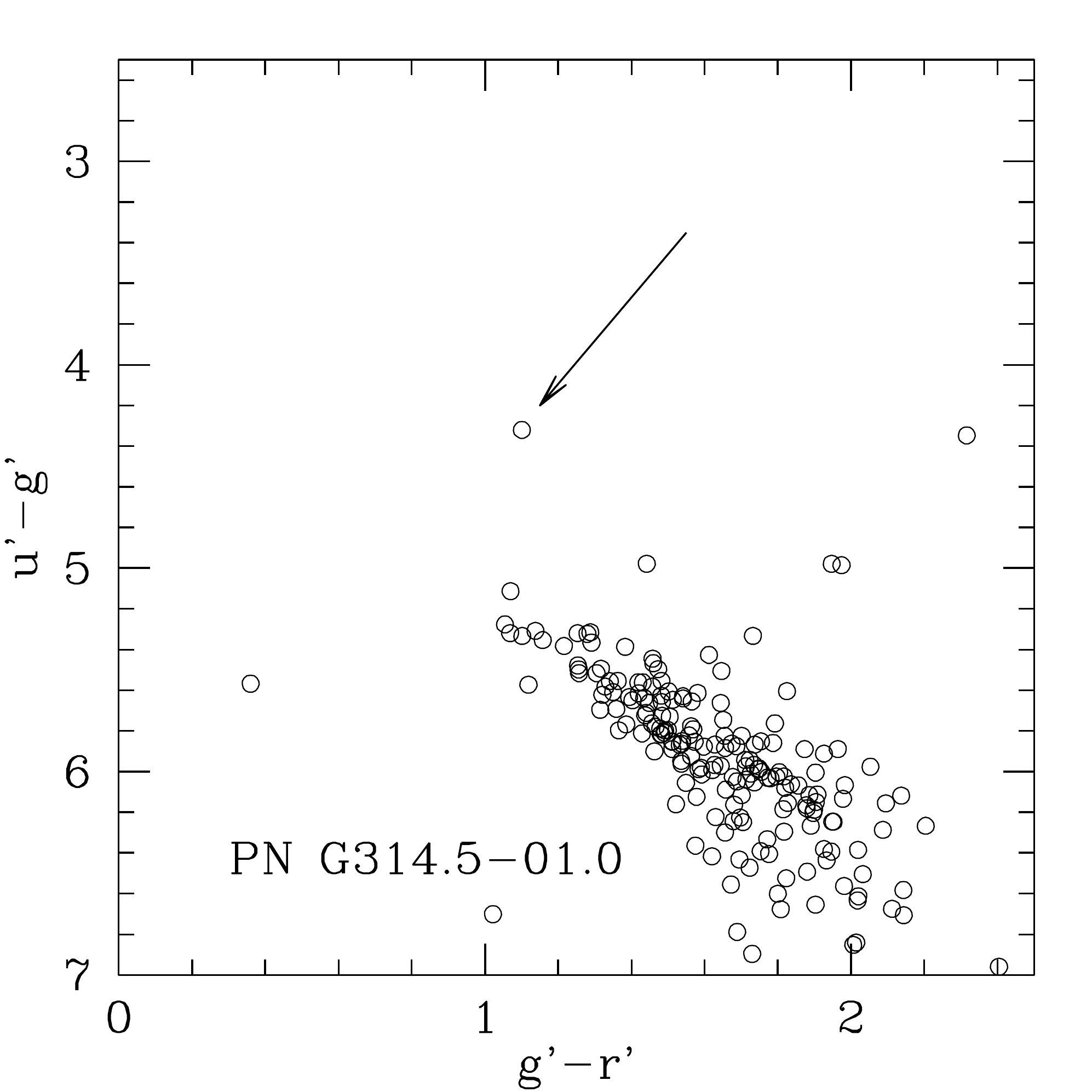}

\caption{Instrumental  $(u'-g')$ vs.\ $(g'-r')$ diagrams of the observed
areas around  the planetary nebulae under study.
Each area is $5.5\times5.5$~arcmin$^2$.
 The proposed central stars
of the nebulae are marked by arrows.\label{fig:c-cd1}}
\end{figure}

\begin{figure} 
\centering
\figurenum{1}
\includegraphics[scale=0.34]{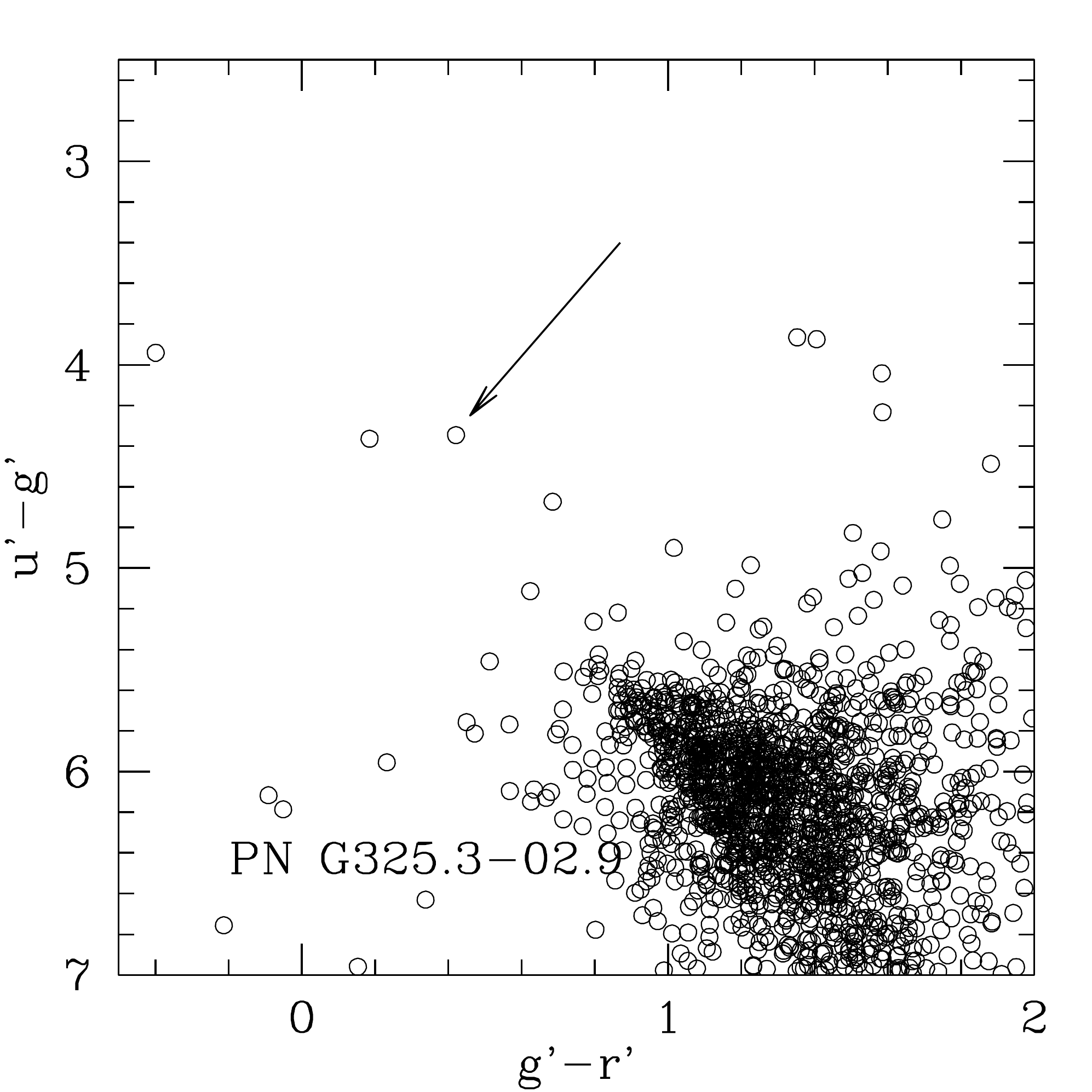}
\includegraphics[scale=0.34]{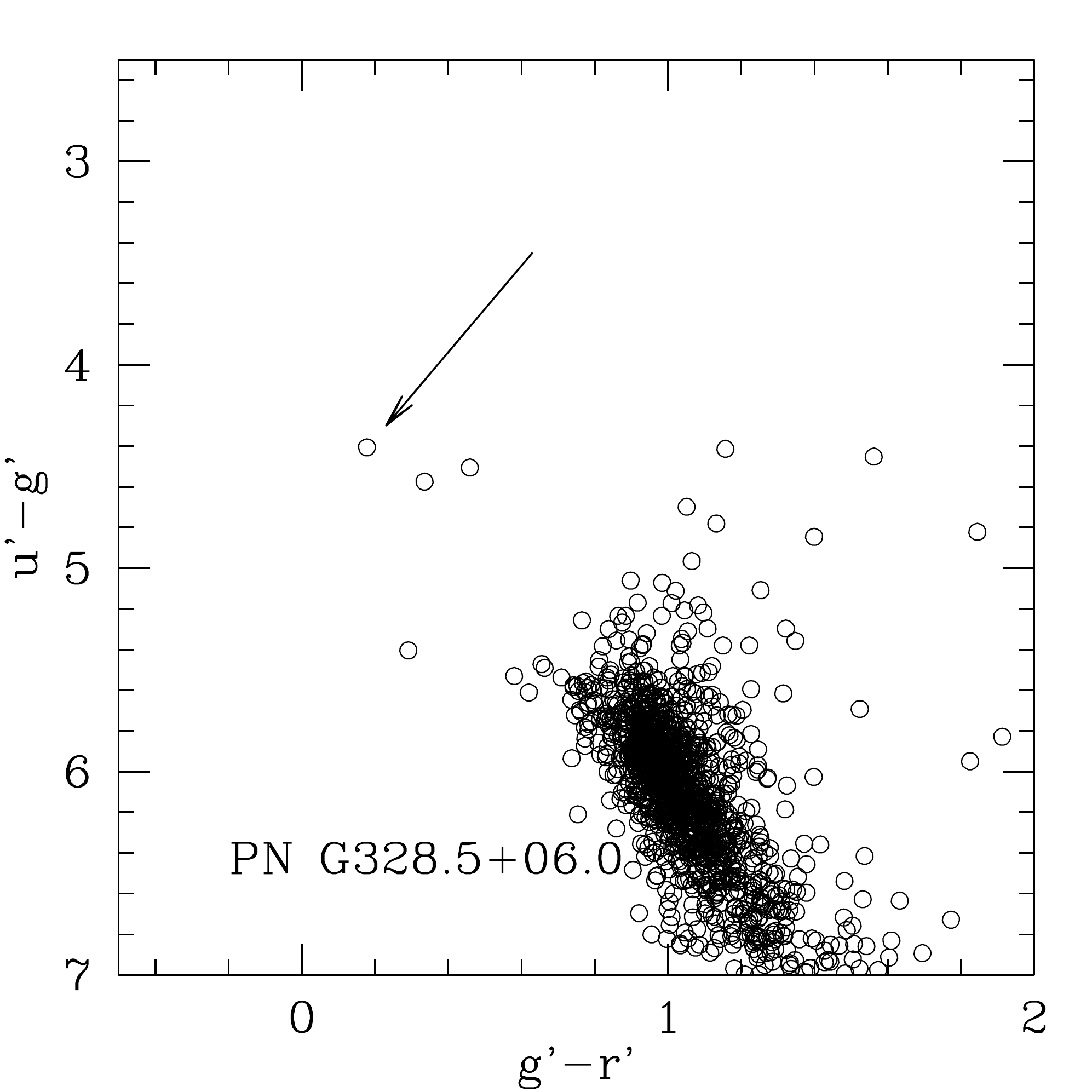}

\includegraphics[scale=0.34]{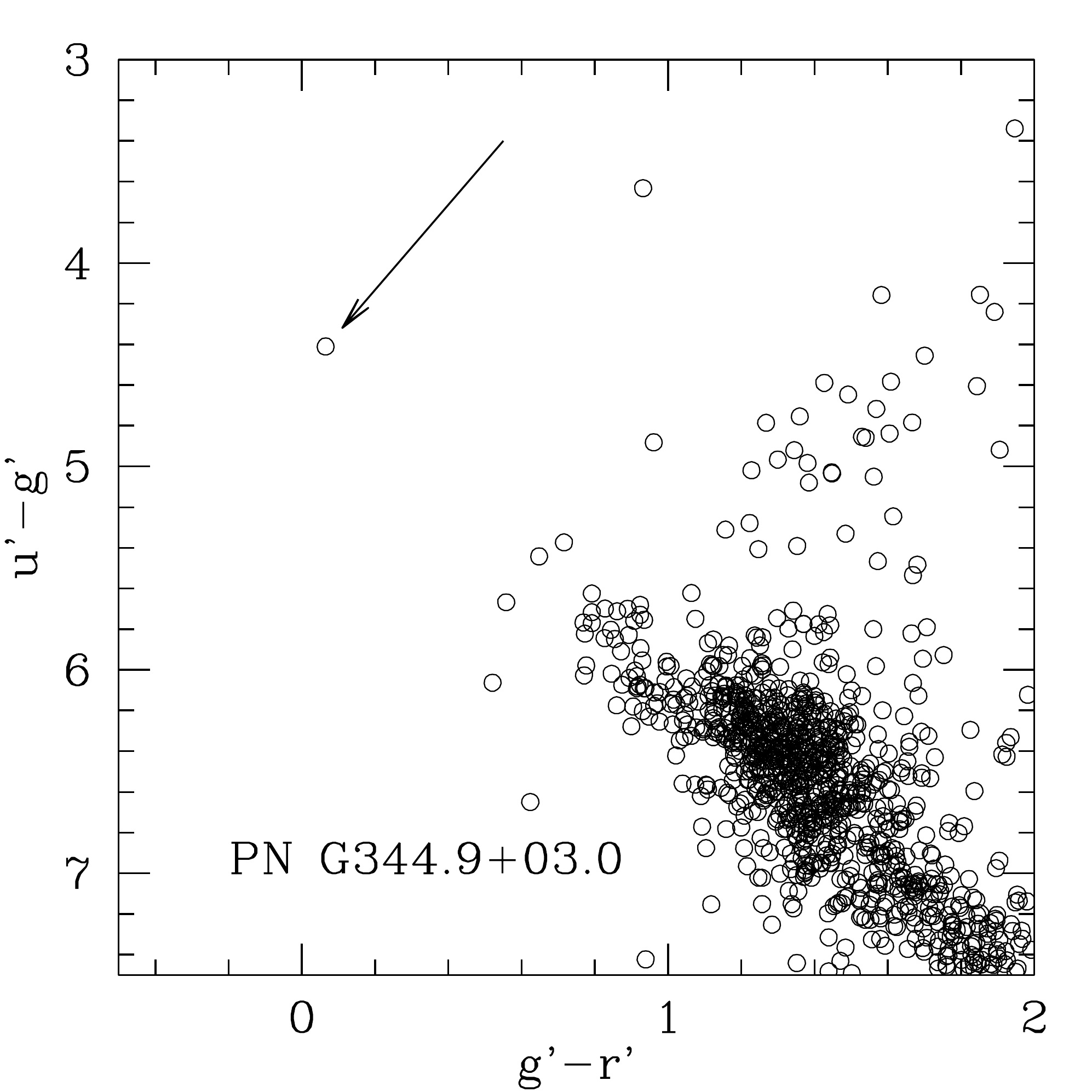}
\caption{Continued. 
\label{fig:c-cd2}}
\end{figure}

To estimate the Sloan magnitudes of the stars we made use of the AAVSO Photometric 
All-Sky Survey (APASS);\footnote{\url{https://www.aavso.org/apass}} 
this is conducted in five filters: 
Johnson $B$ and $V$, and Sloan $g'$, $r'$, and $i'$, and is valid from 
about 7th 
to about 17th magnitude. Starting with some stars in the field with available APASS 
$r'$ magnitude, and comparing these magnitudes with our instrumental values,
we estimated the  $r'$ magnitudes of the CSPNe  listed in Table~\ref{tab:np}.


\section{Spectral classification}
\label{sec:esp}

\subsection{Main features of the spectra}

Fig.~\ref{fig:perfil1} shows 
the Gemini spectra of the stars found at the centers of PN~G019.7$-$10.7, 
PN~G237.0$+$00.7, PN~G276.2$-$06.6, PN~G298.7$-$07.5,
PN~G302.1$+$00.3, and
 PN~G325.3$-$02.9.
They display some evident  features such as
the Balmer series and a couple of {He\,{\footnotesize II}} lines,
all of them fairly broadened.
With this information we are able to say that these stars belong 
to the H-rich group and are very hot.
 
Since Stark effect is  the main cause of the broadening of absorption
lines in spectra of hot WDs \citep{2009ApJ...696.1755T}, 
 the value of the full-width
at half-maximum (FWHM) of spectral features 
could  be a good criterion to help distinguish WDs from
Population~II O-type stars. 
In Table~\ref{tab:esp-wd_b}  we 
put together values of FWHM of several lines for our CSPNe,
 a sample of early O-type Population~II stars and WDs from \cite{2018A&A...614A.135W}, 
 and early~O subdwarfs from 
\cite{2013A&A...551A..31D}. It is evident that the FWHM of absorption lines  of
our  stars  are compatible with those of WDs.
Moreover, the asymmetrical profiles shown by the Balmer
lines could be caused by the presence of {He\,{\footnotesize II}} lines, 
which is typical of WDs spectra 
\citep{1987ApJS...65..603M}.
 Given all this evidence,
 we  classify  the CSPNe of PN~G019.7$-$10.7, 
PN~G237.0$+$00.7, PN~G276.2$-$06.6, and
 PN~G325.3$-$02.9 as WDs of the DAO subtype.

However, we must point out 
that the spectra  of PN~G276.2$-$06.6 and PN~G237.0$+$00.7
also display a combination of emission 
and absorption in H$\alpha$.
This remarkable feature is particularly  evident in the spectrum of PN~G276.2$-$06.6. 
We believe that this emission is real, because the
 method used in Section~\ref{essp} for subtracting the 
 nebular contribution is especially efficient in large objects.
 We note that this characteristic is also present in the hot O-type
 subdwarf BD$+28$~4211 \citep{1999PASP..111.1144H}. 
 Subdwarf O~stars (sdOs) share properties with some CSPNe and DAO-type WDs 
 \citep{2016PASP..128h2001H}; there are few sdOs identified
 as CSPNe, and they are briefly discussed in Section~\ref{sec:subenanas}.

The central star of RCW~69 was first pointed out by 
\cite{2006MNRAS.372.1081F}.
The authors did not obtain spectra of this star, and the physical 
parameters  they give are highly uncertain.
Anyway, we agree with their identification, and
 classify this star as a possible WD.

We were not able to remove the nebular contribution in 
the spectrum of the central star of PN~G325.3$-$02.9;
nevertheless, it shows evident, wide absorption lines of H and \ion{He}{2}.
The spectrum of the central star of PN~G298.7$-$07.5
    is very noisy and we could not detect any line of \ion{He}{2} in it, so
for now we  classify the object as a possible WD.
Finally, the spectrum of the central star of
PN~G314.5$-$01.0 has a moderate S/N but  it does not display any features.
Note that this  does not necessarily rule out a WD classification;
we  categorize this star as ``continuum''
\citep{2018A&A...614A.135W}.

\begin{figure}
\includegraphics[scale=1.16]{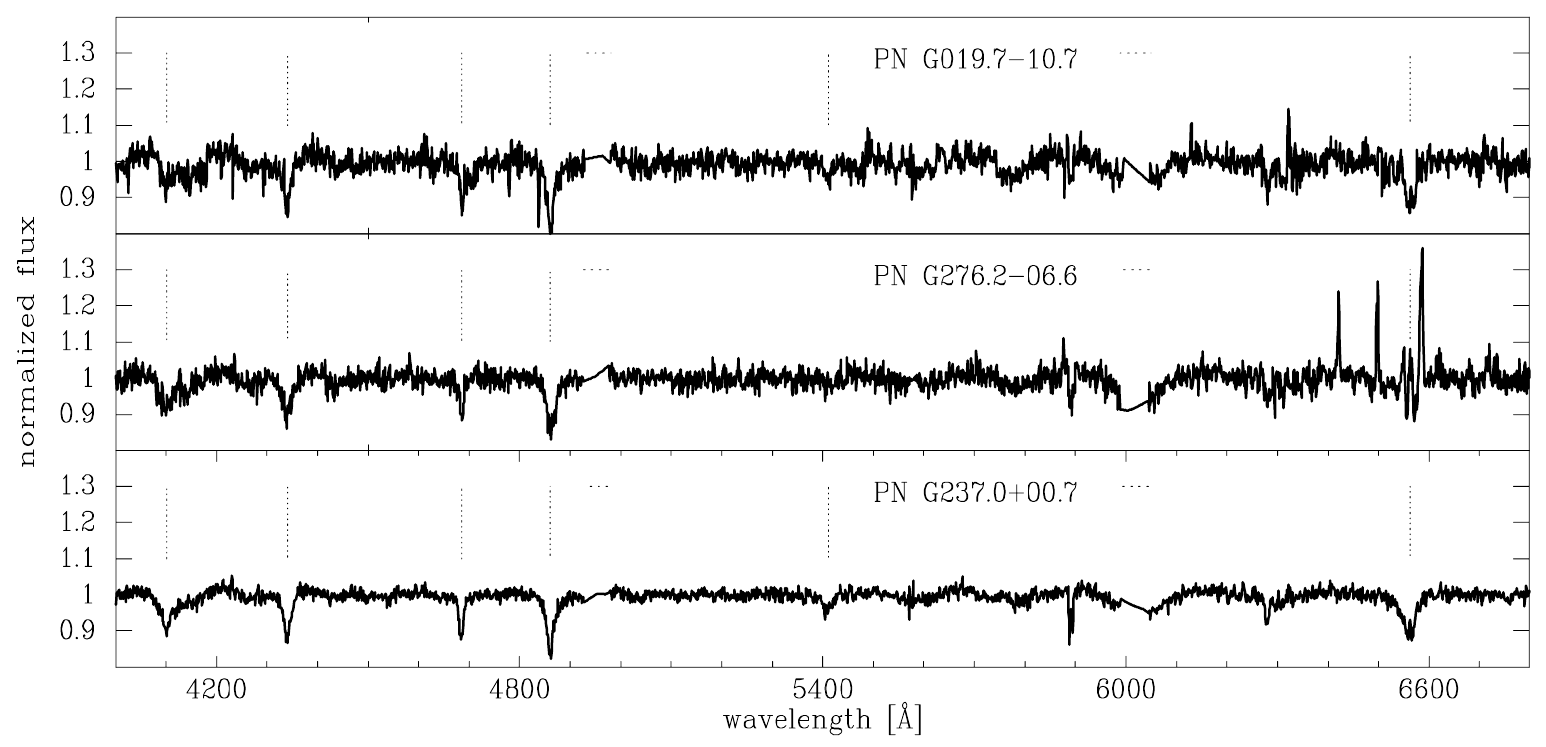}
\caption{Normalized Gemini spectra of the central stars of 
PNe of our sample.
  The Balmer lines H$\alpha$, H$\beta$, H$\gamma$, 
and H$\delta$, and the \ion{He}{2} lines at
      4686~\AA\ and 5412~\AA\ 
      are indicated with vertical dotted lines.
      The interstellar  D-lines of \ion{Na}{1}  at 5890 and 5896~\AA\ are not marked, but  are
      clearly seen in the spectra. It is  noteworthy
      the absorption line at 6279~\AA, a feature  of unknown origin that  also appears in the 
      spectrum of the O(He)-type central star of the nebula \object[PN K 1-27]{K~1-27}
      \citep{1996A&A...310..613R, 1998A&A...338..651R}.
    The horizontal dotted lines mark the positions
      of the gaps separating the CCDs of the spectrograph.}
\label{fig:perfil1}
\end{figure}

\begin{figure}
\includegraphics[scale=1.16]{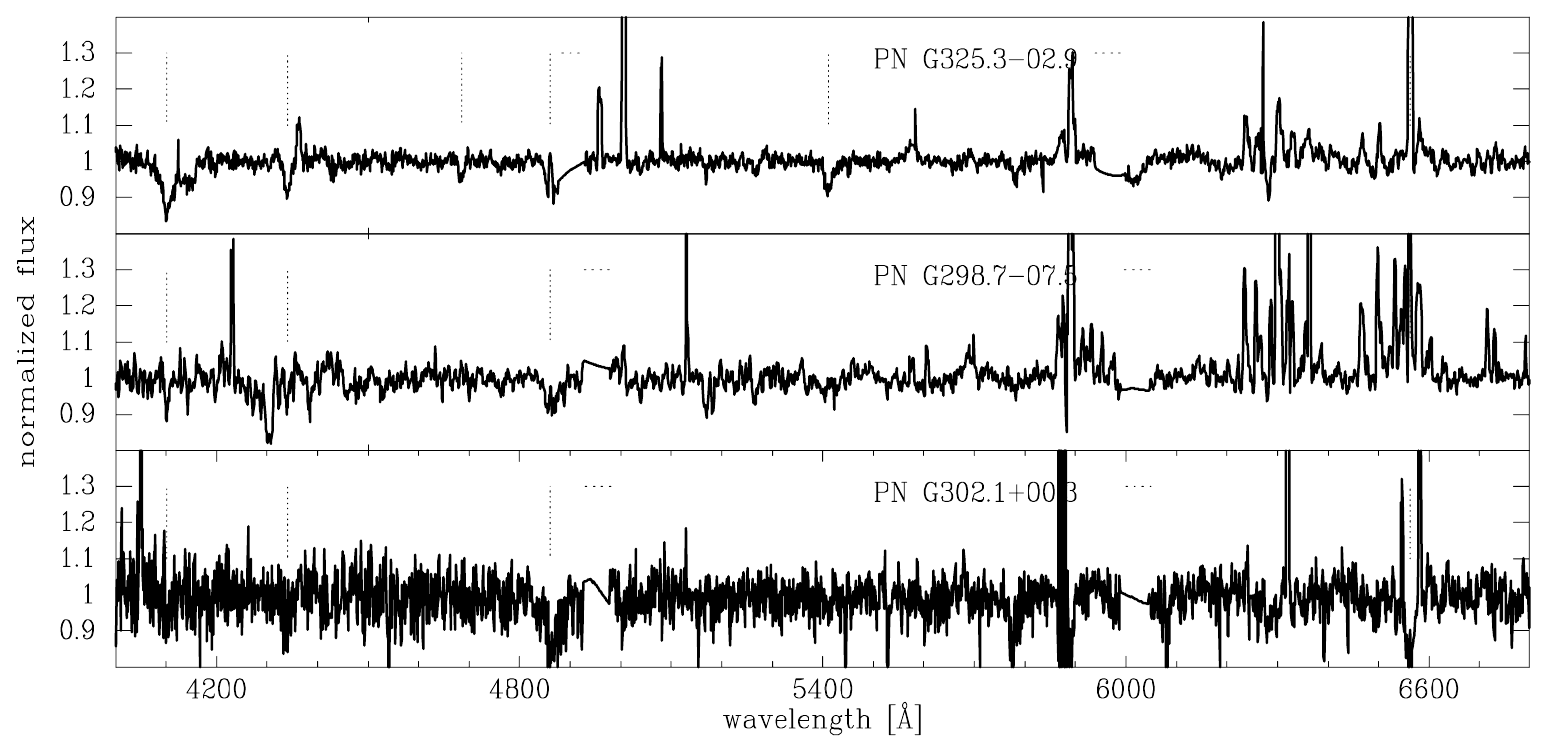}
\figurenum{2}
\caption{Continued.
 } 
\label{fig:perfil2}
\end{figure}

\begin{deluxetable*}{lccccc} 
\tablecaption{Average full-width at half-maximum (FWHM, in \AA)
of absorption lines 
\label{tab:esp-wd_b}}
\tablewidth{0pt}
\tablehead{
\colhead{Object} &
\colhead{S/N\tablenotemark{a}} & 
\colhead{FWHM(H$\gamma$)}  &
\colhead{FWHM(4686)} &
\colhead{FWHM(H$\beta$)} &
\colhead{$N$\tablenotemark{b}}  
} 
\startdata
PN~G019.7$-$10.7                  &  46    & 14      & 10      & 20      & \nodata       \\
PN~G237.0$+$00.7                  & 120    & 18      & 11      & 20      & \nodata       \\
PN~G276.2$-$06.6                  &  60    & 25      &  9      & 23      & \nodata       \\
PN~G298.7$-$07.5                  &  32    & \nodata & \nodata & 31      & \nodata       \\
PN~G302.1$+$00.3                  & 18     & \nodata & \nodata & 37      & \nodata       \\ 
PN~G314.5$-$01.0                  &  30    & \nodata & \nodata & \nodata & \nodata       \\
PN~G325.3$-$02.9                  & 65     & 14      & 10      & \nodata & \nodata       \\
%
O-type Pop.~II stars\tablenotemark{c}  & \nodata   & $9\pm2$    & $6\pm2$  & $9\pm2$   & 12   \\
sdO2--3\tablenotemark{d}               & \nodata   & $13\pm3$   & $8\pm2$  & $13\pm2$  &  8   \\
white dwarfs\tablenotemark{c}          & \nodata   & $>13$      & $>10$    & $>17$     &  3   \\
\enddata
\tablenotetext{a}{Ratios measured in the range  5050--5200~\AA.}
\tablenotetext{b}{Number of objects used in the statistics.}
 \tablenotetext{c}{Measurements on good-quality spectra of
 early O-type stars and WDs presented  by Weidmann et al.\ (2018).}
\tablenotetext{d}{Measurements on spectra published by Drilling et al.\ (2013), 
considering only the hottest subtypes.}
\end{deluxetable*}


\subsection{Effective temperatures and surface gravities   derived with TheoSSA} 
\label{sec:modelo}

To derive the stellar parameters $T_{\mathrm{eff}}$ and $\log g$  from our spectra,
we employed the TheoSSA\footnote{\url{http://astro.uni-tuebingen.de/~TMAW}} model,
which is especially suitable for hot and compact stars like WDs.
We followed the recipe given by \cite{2018MNRAS.475.3896R} to properly
 fit  the line profiles. 
We considered  stellar atmospheres composed only of H+He
with solar abundances.

To address the problem reported by
\cite{1994A&A...285..603N}, namely, that
very different temperatures may be derived from
the fits of different Balmer lines,
 we adopted the criterion
of fitting  the H$\delta$ line.
The only exception was
 PN~G019.7$-$10.7, whose H$\delta$ is very noisy; 
 in this case,
we used H$\gamma$ instead.

 Concerning the S/N ratio of our spectra, 
we want to remark that they were normalized and corrected by radial velocity.
This  procedure is essential to achieve a meaningful  comparison with the TheoSSA model but, 
 unfortunately, 
 it degrades even more the quality of the data.
It is difficult to evaluate the uncertainties of the derived parameters, 
given  the low quality of our spectra. 
Rough estimates of the uncertainties might be:   $\Delta T_{\mathrm{eff}}=7000$~K
and  $\Delta \log g=0.3$.

Our results for four CSPNe are displayed in Fig.~\ref{fig:theo1}.
 The star of
PN~G325.3$-$02.9 deserves an extra comment. 
Its spectrum shows line profiles that are distorted, probably  
due to the normalization operation.
The  solution that we could obtain entails a couple of parameters that
do not correspond to a CSPN.
Here we prefer to adopt uncertainties that at least
double those given above.

Finally, we could not perform  good fits for the objects PN~G298.7$-$07.5 and
PN~G302.1$+$00.3, because their spectra are very noisy.

\newpage

\begin{figure}
\includegraphics[scale=0.45]{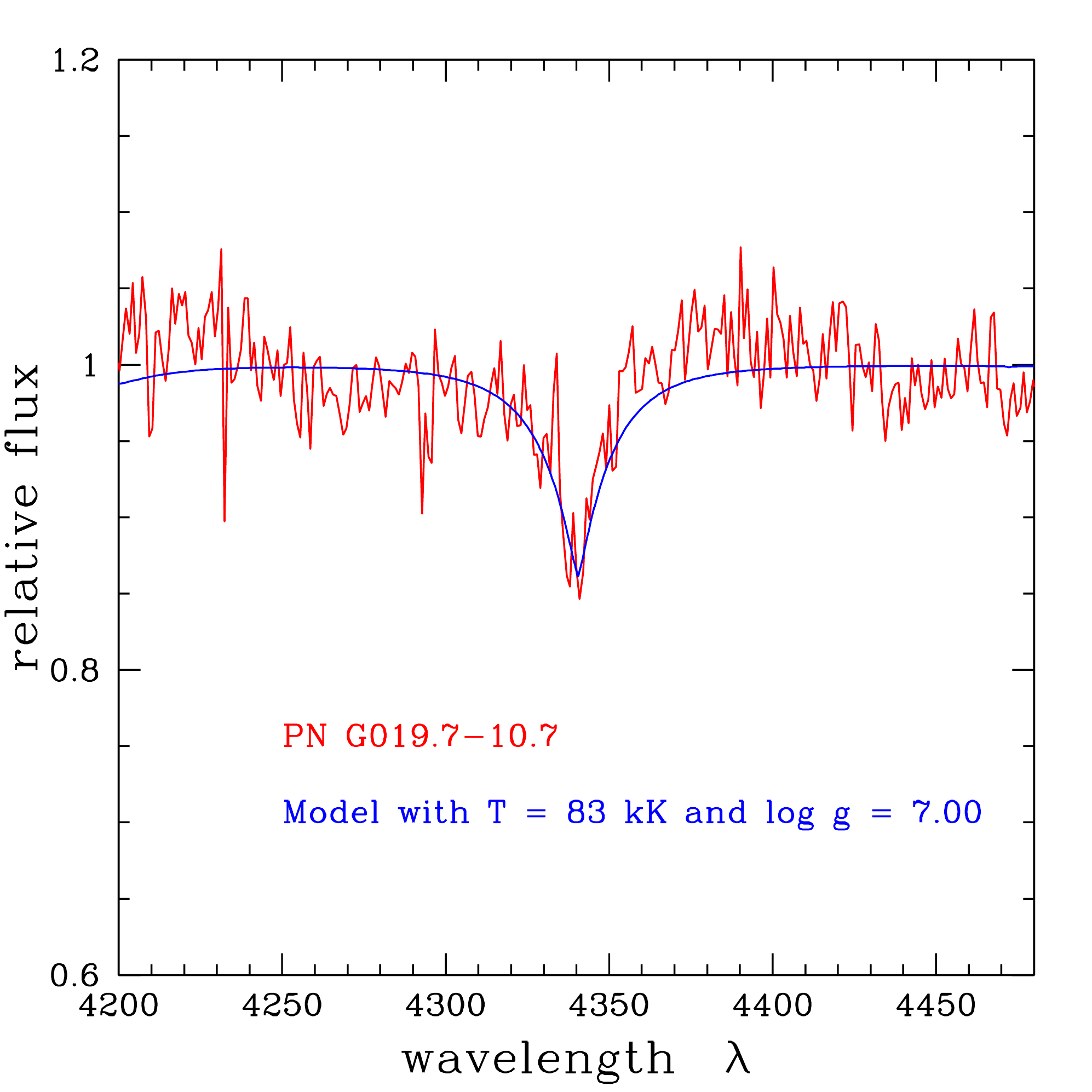}
\includegraphics[scale=0.45]{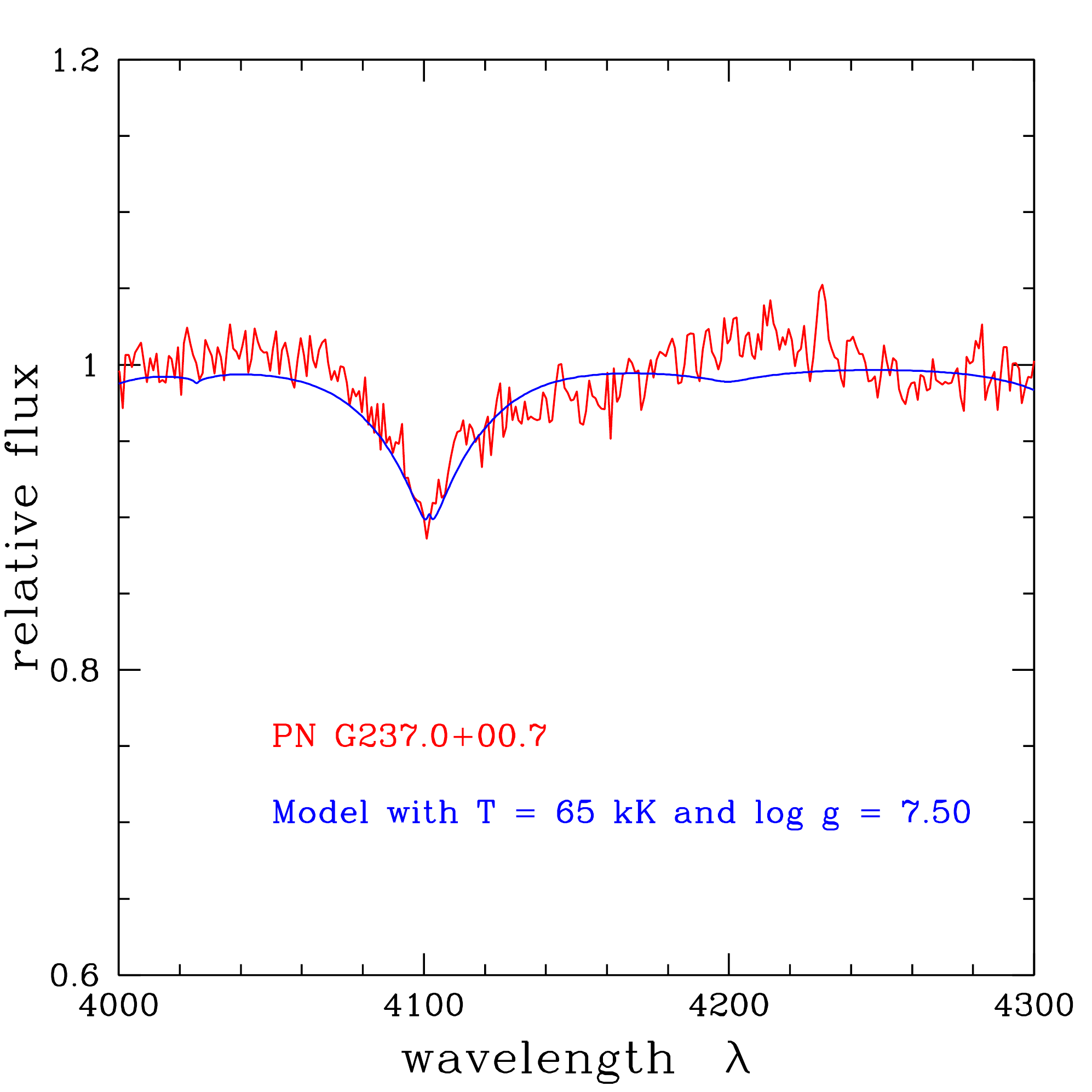}
\includegraphics[scale=0.45]{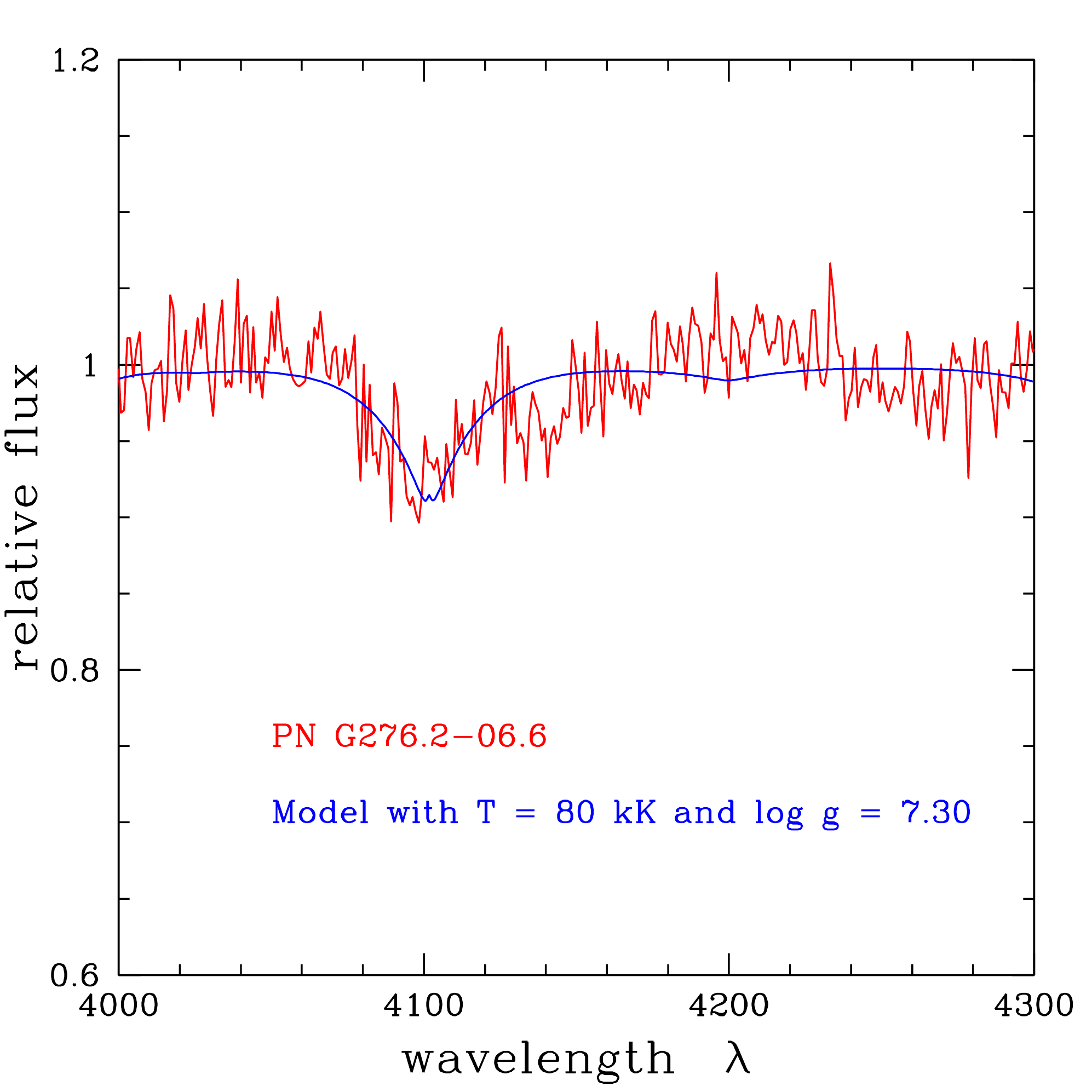}
\includegraphics[scale=0.45]{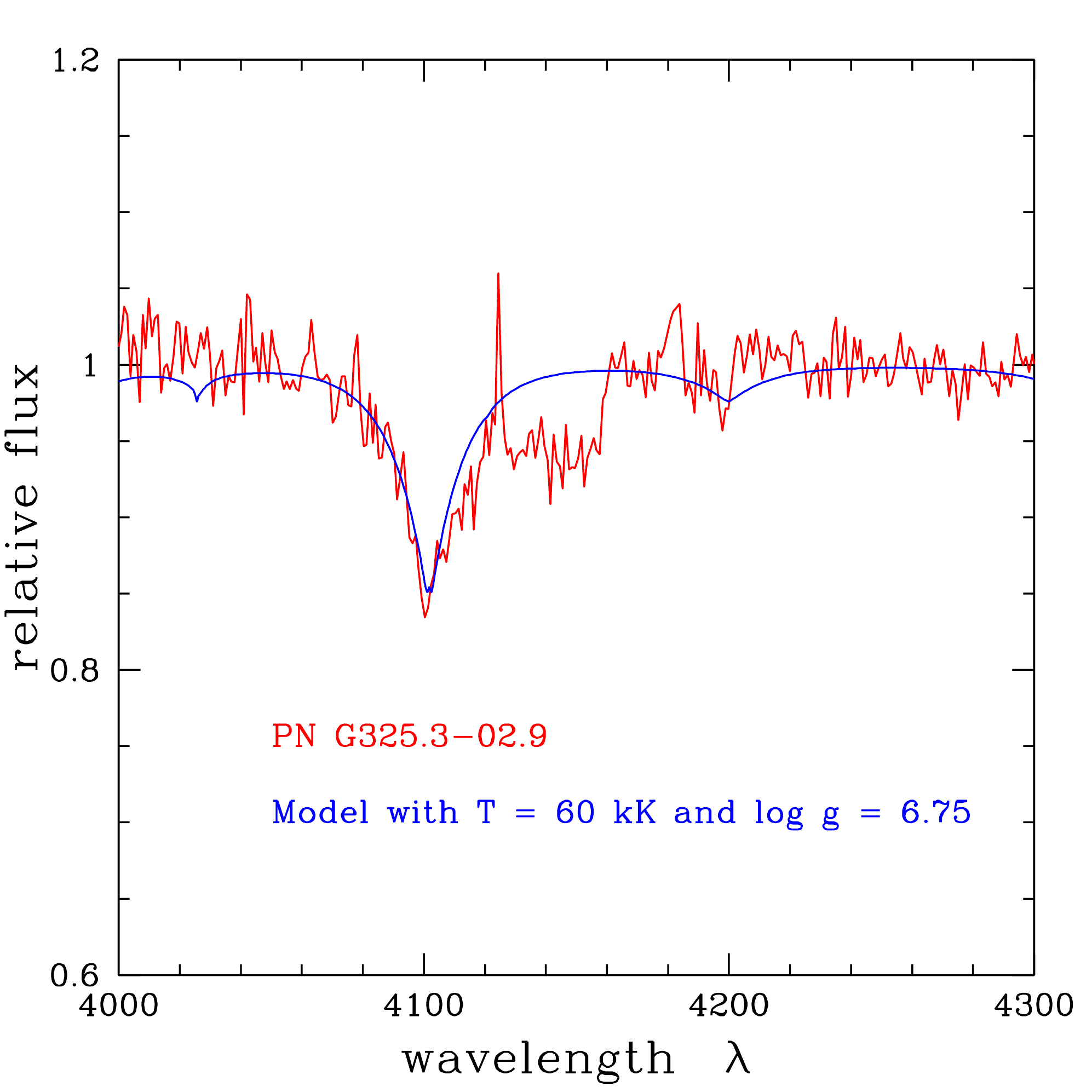}
\caption{Our line profiles compared with synthetic TheoSSA spectra. 
The line fit is H$\delta$, except 
for PN~G019.7$-$10.7, where 
  H$\gamma$ was used. The units of $g$ are cm~s$^{-2}$.
  } 
\label{fig:theo1}
\end{figure}

\begin{figure}
\includegraphics[scale=0.45]{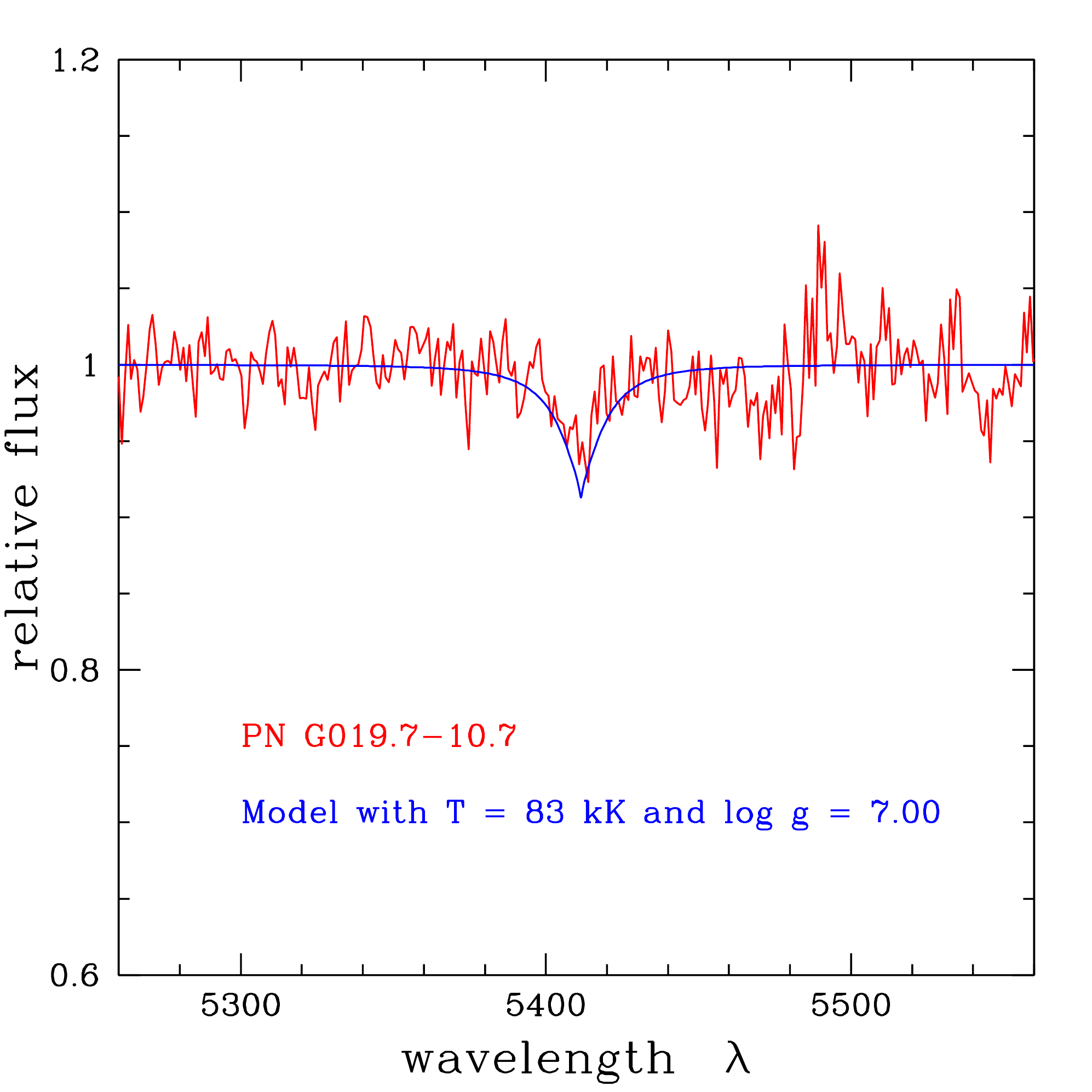}
\includegraphics[scale=0.45]{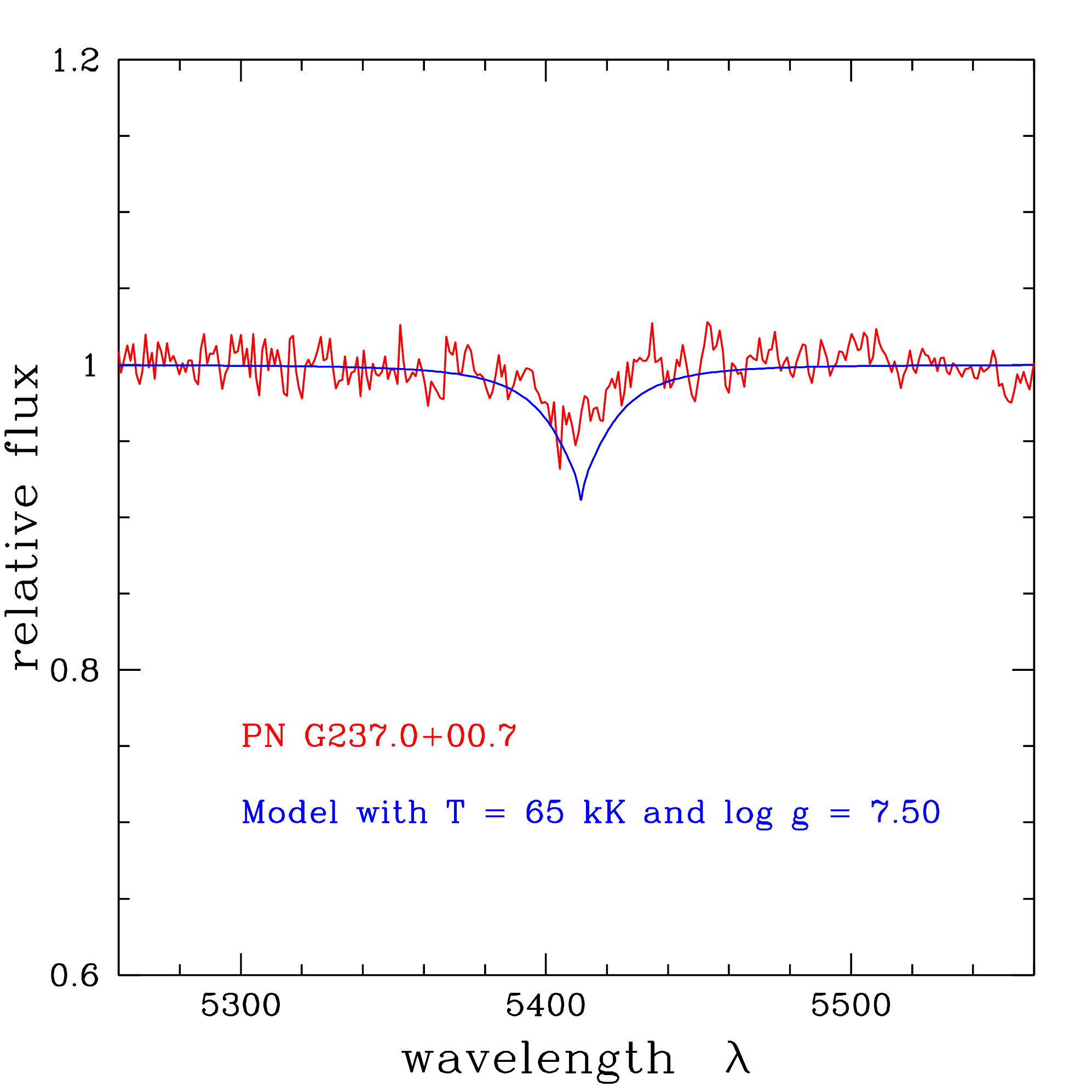}
\includegraphics[scale=0.45]{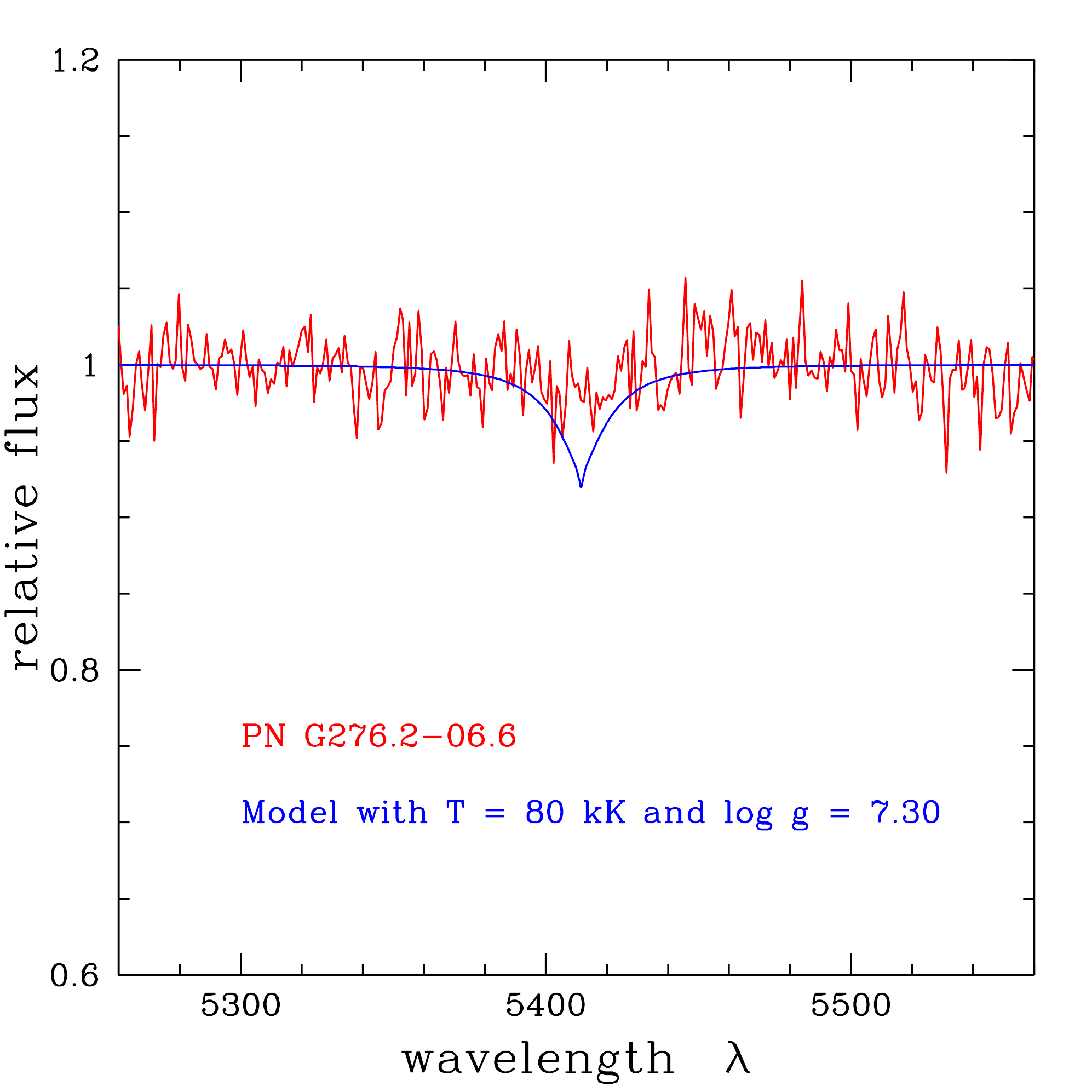}
\includegraphics[scale=0.45]{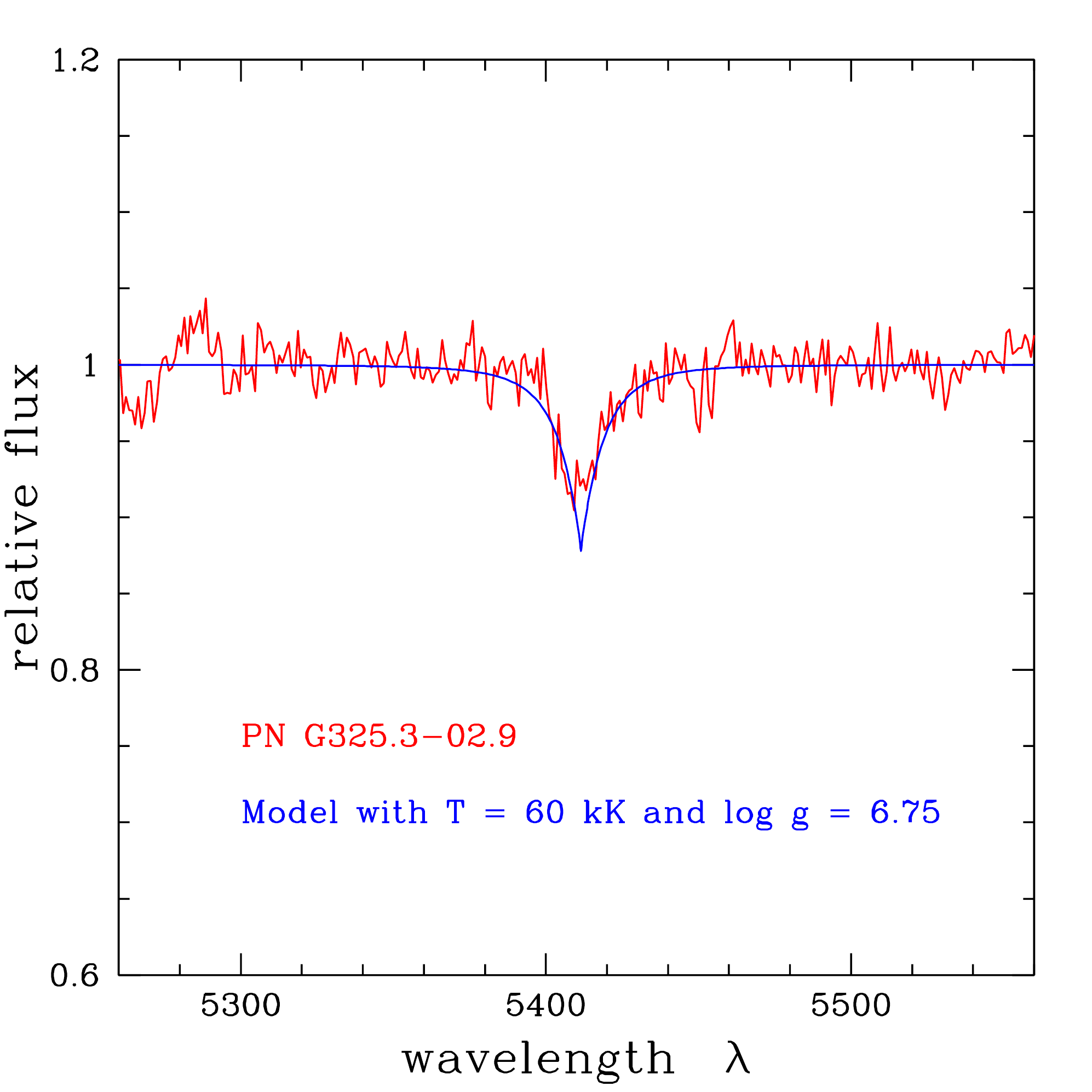}
\figurenum{3}
\caption{Continued. 
The \ion{He}{2} (5411~\AA) line is shown.}
\label{fig:theo2}
\end{figure}

\newpage


\subsection{White dwarfs or hot subdwarfs as CSPNe?}
\label{sec:subenanas}

Although  sdOs are relatively common objects,
they appear  scarcely related to PNe.
   Attempts to find nebulae around sdOs have had little success   so far
   (\citealt{1988A&A...198..287M},
\citealt{1989AJ.....97.1423K}), so much so that
 only five sdOs are currently known to be  nuclei of PNe
(\citealt{2002A&A...381.1007R},
\citealt{2013A&A...552A..25A},
\citealt{2015MNRAS.446..317A}).
Of these stars perhaps the most relevant one is the sdO that ionizes
the \object[PN Ps 1]{PN~Ps~1}, since this nebula appears physically associated to the
globular cluster \object{M15}.  
 There are four known PNe associated with globular clusters:
 Ps~1 in M15, \object[IRAS 18333$-$2357]{IRAS~18333$-$2357} 
 in \object{M22}, \object[PN JaFu 1]{JaFu~1} in \object[Pal 6]{Pal~6}, 
 and \object[PN JaFu 2]{JaFu~2} in \object[NGC 6441]{NGC~6441}, with
  progenitor masses that range from 0.8 to 1.2~$M_\odot$
 \citep{2017ApJ...836...93J}.
 Thus far only for the first two  nebulae it has been possible
to assess the spectral type of their CSPNe, being sdO
 \citep{we11}.  This suggests 
that  sdO is a final stage for stars of very low mass.

Since sdOs belong to the H-rich  group of CSPNe
\citep{2013A&A...551A..31D} like the DAs,
 how to distinguish one from another?
According to \citet{2016MNRAS.455.3413K},
 sdOs have  $\log g < 6.5$  and  DAs have  $\log g > 7.0$.
 Consequently, and given the results of the current Section, we can
 safely assume that
 our objects are indeed WDs.


 \newpage

 \section{Properties of the CSPN derived from GAIA  DR2 
data\label{sec:gaia}}

\subsection{Magnitudes and colors}

\begin{figure}
\centering
\includegraphics[scale=0.4]{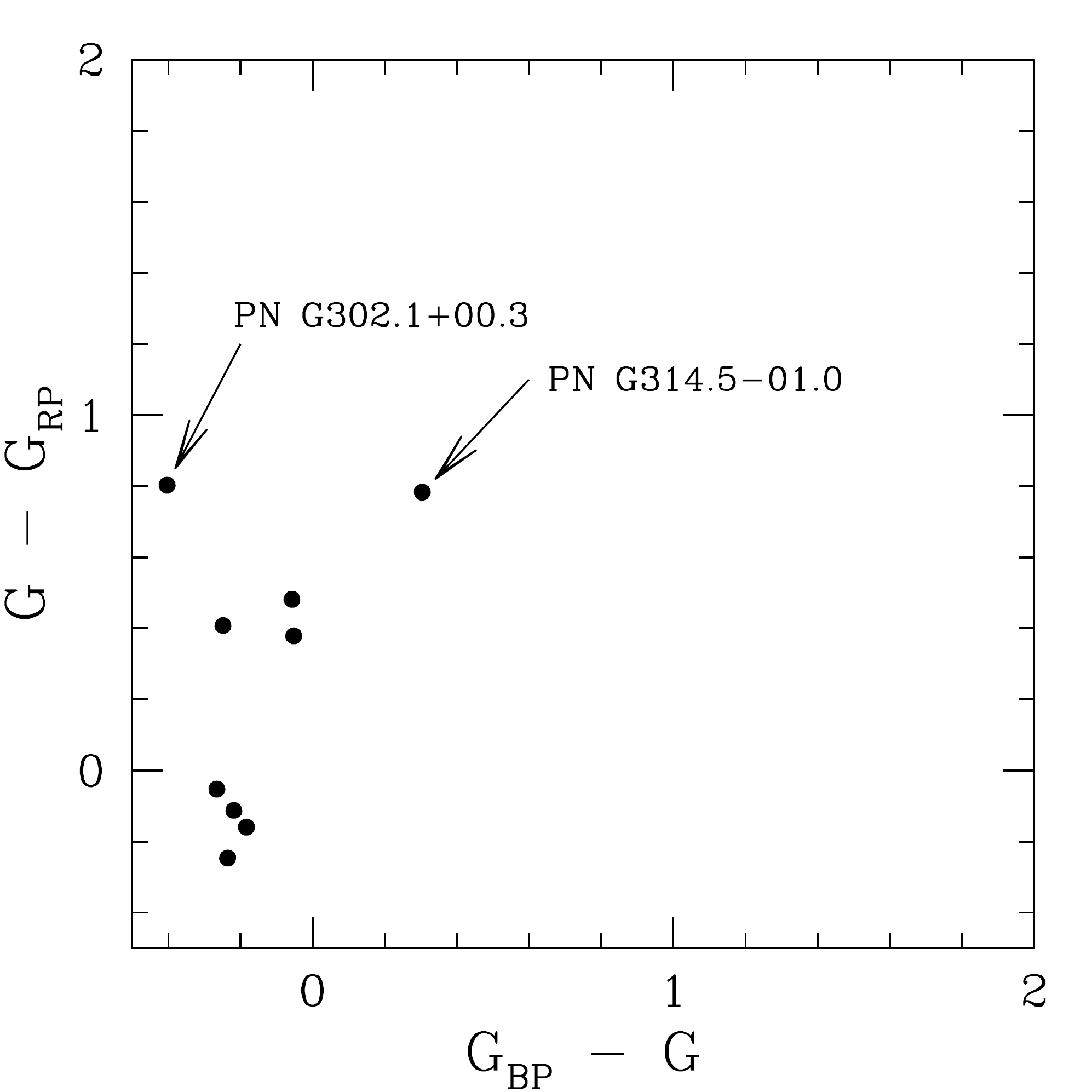}
\caption{Color-color diagram in the \emph{Gaia} 
photometric system for nine CSPNe.}
\label{fig:enrojecimientos}
\end{figure}

The CSPNe under study were searched  in the \emph{Gaia} Data Release~2
Archive\footnote{\url{https://gea.esac.esa.int/archive/}} 
\citep{gaia16,gaia18} for  information on  parallaxes
and colors, to help us characterize them better.
The results of the search are in Table~\ref{tab:gaia}.
It can be verified that \emph{Gaia}'s $G$~magnitudes and
our APASS-estimated $r'$~magnitudes (Table~\ref{tab:np}) 
are fairly similar. Although our sample is composed of early-type
stars, some colors appear positive, especially those of 
{PN~G314.5$-$01.0, (whose colors, however,  are consistent} with
the star's position in our color-color diagram in Fig.~\ref{fig:c-cd2}).
According to \cite{and18},
indices ($G_\mathrm{BP} - G$) and ($G- G_\mathrm{RP}$) are
independent of the distance, unlike $(G_\mathrm{BP} - G_\mathrm{RP})$.
{Therefore,} positive values of the said indices  imply considerable reddenings.
Assuming that these colors are correct, we can plot
 the  ($G_\mathrm{BP} - G$)
vs.\ ($G- G_\mathrm{RP}$) diagram (Fig.~\ref{fig:enrojecimientos}).
In this Figure most points 
follow roughly a line, in which the most reddened star---although not
the farthest one, see   Section~\ref{sec:distancias}---is indeed that of PNG~G314.5$-$01.0.
For comparison, a similar  plot {for dereddened stars} like  Figure~4 of \cite{and18},
indicates that our stars, that must be of approximately the same intrinsic color,
should be grouped towards the lower left corner.
In Fig.~\ref{fig:enrojecimientos} the central star of PN~G302.1$+$00.3 seems to
be the exception to the trend:  perhaps its 
reddening is  altered by the absorption originated in
the nebula itself; in fact, its finding chart (Fig.~\ref{fig:cartas}~(\emph{e}))
{is the only one 
that shows clearly the nebula,} at least through the
  filter $r'$.

\subsection{Distances}
\label{sec:distancias}

The use of the estimator ``parallax'' ($\varpi$), given with an uncertainty $\sigma_\varpi$, to \emph{estimate}
the distance $r$ has been  thoroughly  discussed by \cite{luri18} for the \emph{Gaia} DR2 data. They
show that the naive, direct interpretation of the distance as the simple inverse of the parallax 
is only accurate when the relative error $f = \sigma_\varpi / \varpi$  is at most   20\%. Larger 
values of $f$ require another, more careful approach, that should take advantage of the information 
that those imprecise  (even \emph{negative}, see Table~\ref{tab:gaia}) parallaxes may carry. 
\cite{luri18} recommend  tackling
this problem as an inference {one,} to be  preferably handled
with a full Bayesian approach. This method has been treated in depth by \cite{bailer15}: {it}
  involves estimating a ``posterior probability''
  $P(r|\varpi,\sigma_\varpi)$ 
over $r$, given the observables ($\varpi$, $\sigma_\varpi$), as follows:
\begin{equation}
    P(r|\varpi,\sigma_\varpi) = \frac{1}{Z}
    P(\varpi|r,\sigma_\varpi)P(r).
\end{equation}
This is Bayes' Theorem, in which $P(\varpi|r,\sigma_\varpi)$ is the conditional probability 
of the observable parallax $\varpi$ given $r$
and $\sigma_\varpi$, $P(r)$ is the prior probability (or simply ``prior''), and $Z$ 
is a normalization constant.
The estimate of the distance is then the \emph{mode} of the pdf $P(r|\varpi,\sigma_\varpi)$.
For the measurement model used in the  {\it Gaia} data processing \citep{bailer15}:
\begin{equation}
P(\varpi|r,\sigma_\varpi) = \frac{1}{\sqrt{2\pi}\sigma_\varpi} 
\exp\left[- \frac{1}{2\sigma^2_\varpi}\left(\varpi - \frac{1}{r}\right)^2\right].
\end{equation}
The prior expresses our knowledge of---or assumptions about---the distance, 
independent of $\varpi$.
\cite{bailer15} and \cite{astraat1,astraat2}
discuss  several priors, among them  the
``exponentially decreasing space density:''
\begin{equation}
    P(r) = \left\{ \begin{array}{lll}
            \frac{1}{2L^3}\, r^2 e^{-r/L} & & \mathrm{if} \,\, r>0, \\
            0 & &\mathrm{otherwise,}
            \end{array} \right.
\end{equation}
with $L$ a length scale.
We choose it to estimate the distances of our CSPNe because it is {simple, 
reasonable, and  is available in}  the
\textsc{TopCat}\footnote{\url{http://www.starlink.ac.uk/topcat/}} software. 
The distances are in Table~\ref{tab:gaia}, given with 90\% confidence intervals 
as recommended by \cite{bailer15}.
We took $L=1.35$~kpc, value suggested in the cited papers.
The estimated  distances appear to be reasonable, and are
even consistent with the scale of relative apparent magnitudes,
with the exception of
 PN~G325.3$-$02.9.  The \emph{Gaia} parallax for this star is markedly negative, 
which results, through  the adopted model, in
an exceedingly large distance, in conflict  with its  brightness (Table~\ref{tab:gaia})
and   reddening (Fig.~\ref{fig:enrojecimientos}).

\begin{deluxetable*}{ccccccccc} 
\tablecaption{\emph{Gaia} data and derived distances for the central stars
of our sample of PNe
\label{tab:gaia}}
 \tablewidth{0pt}
\tablehead{%
\colhead{PN~G} &
\colhead{Gaia DR2 Source Id.} & 
\colhead{Parallax} &
\colhead{\emph{G}} &
\colhead{$G_\mathrm{BP} - G_\mathrm{RP}$} &
\colhead{$G_\mathrm{BP} - G$}  &
\colhead{$G - G_\mathrm{RP}$} &
\colhead{Distance} &
\colhead{90\% CI} \\
 & & (mas) & (mag) & (mag) & (mag) & (mag) & (pc) & (pc)
}
\startdata
019.7$-$10.7  & 4088731114003376512 & $+0.60 \pm 0.21$  & 18.452  & $-0.481$ & $-0.235$& $-0.246$ & 1826.80 & [1324.88, 5622.94] \\
237.0$+$00.7 & 5715387335262528640   & $+0.72 \pm 0.14$ & 18.128 & $-0.342$ & $-0.183$ & $-0.159$ & 1433.15 & [1139.40, 2547.80]\\
276.2$-$06.6  & 5303880196458455808 & $+0.74 \pm 0.19$  & 18.762  & $-0.317$ &$-0.265$ & $-0.052$ & 1453.84 & [1104.38, 3805.05]\\
298.7$-$07.5  &  5855902039382099968 & $-$0.65 $\pm$ 0.51 & 19.973 & $-0.330$ &  $-0.218$ & $-0.112$ & 3903.66 & [2176.32, 9692.69]\\
302.1$+$00.3 & 6055200341668022400 & $+0.40\pm0.33$ & 19.407 & $+0.401$ & $-0.403$ & $+0.803$ & 2603.64 & [1563.92, 8112.72]\\
 314.5$-$01.0  &  5878330972774696704  & $+0.55\pm0.21$ & 18.528 & $+1.087$ & $+0.304$ & $+0.783$ & 1996.02& [1412.34, 6271.82]\\
325.3$-$02.9  & 5835851723298840576  & $-1.14 \pm0.14$ & 16.690  & $+0.424$ & $-0.057$ & $+0.482$ & \nodata & \nodata\\
328.5$+$06.0 & 5986526735199975552   & $+0.66\pm0.38$   & 18.698  & $+0.161$ & $-0.248$  & $+0.408$ & 1981.30 & [1225.92,  7477.57]\\
344.9$+$03.0 & 5970024611844351616 & $+0.51\pm0.31$ &19.027  & $+0.328$ & $-0.052$ & $+0.379$ & 2260.15 & [1431.83, 7583.31]\\
\enddata
\end{deluxetable*}


\section{Comparison with stellar evolution models}
\label{sec:evolucion}

\begin{figure}
\centering
\includegraphics[scale=0.8]{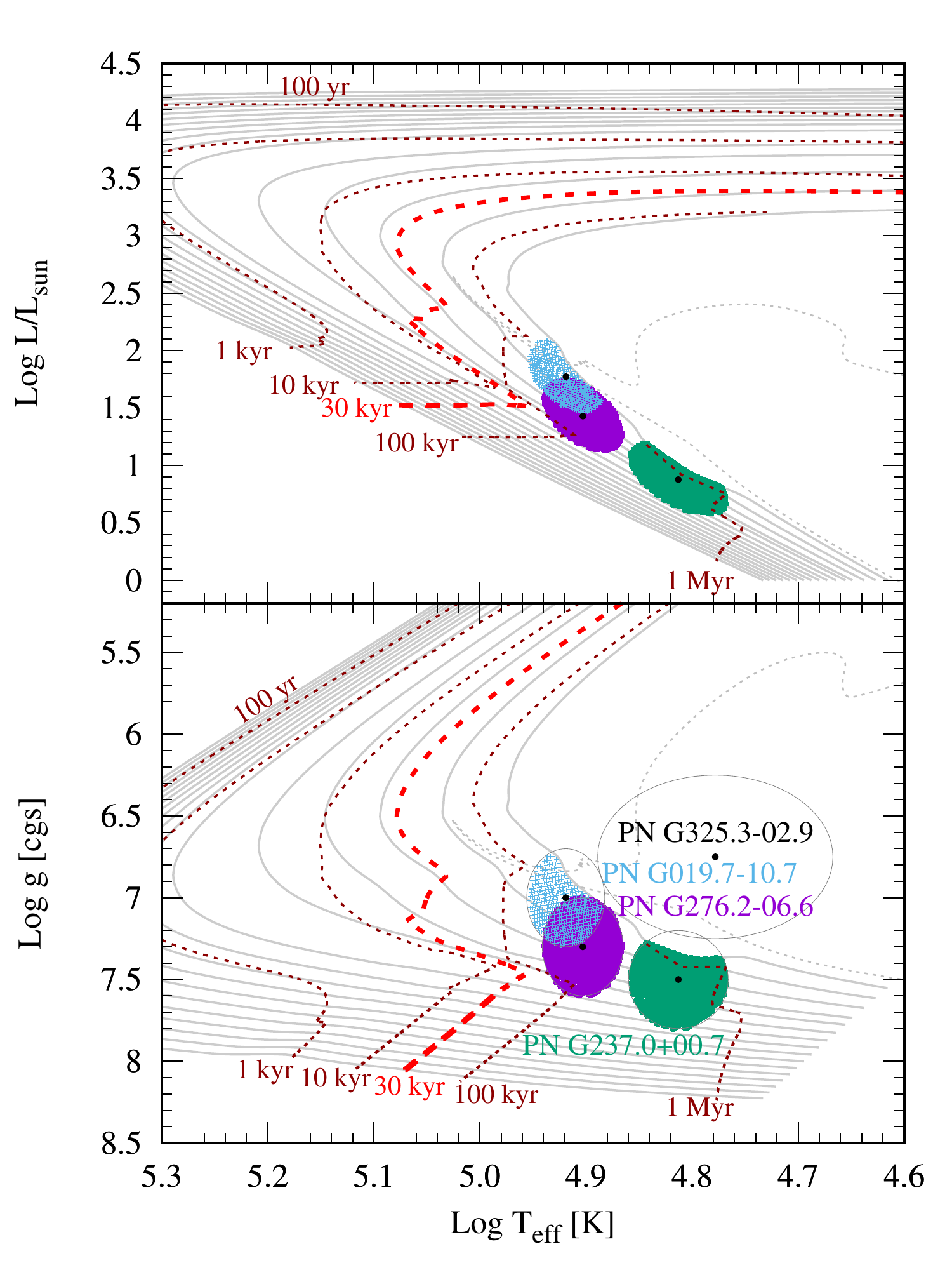}
\caption{\emph{Bottom panel}: Kiel diagram showing the location of the CSPNe with derived 
$\log T_{\rm eff}$ and $\log g$ values, together with tracks interpolated/extrapolated 
from the $Z=0.01$ post-AGB evolutionary sequences of \citet{mm16}. From right to left,
tracks are shown as solid lines for masses from 0.50$M_{\odot}$ to 0.80$M_{\odot}$, with a step of 
0.02$M_{\odot}$. The dashed grey line at low temperatures and gravities 
corresponds to a post hot-subdwarf model ($M=0.4754 M_\odot$) that avoided the AGB phase.  
\emph{Upper panel}: HR diagram showing the luminosities 
of the central stars derived from interpolation/extrapolation from stellar evolution models.}
\label{fig:Kiel_HR}
\end{figure}

For the four CSPNe for which we have values of $T_{\rm eff}$ and $\log g$,  it is 
now possible to derive other stellar properties {using} stellar evolution 
model sequences. In Table~\ref{tab:CSPNe-derived} we show the values of luminosity, age, 
and mass derived for our CSPNe by interpolating/extrapolating in $T_{\rm eff}$ and $\log g$ 
within the $Z=0.01$ post-AGB sequences of \citet{mm16}. The comparison of 
{the parameters} of our CSPNe with the interpolated tracks is shown in the lower 
panel of Fig.~\ref{fig:Kiel_HR}. It is clear from {this Figure} {that,} with the 
current values of  $T_{\rm eff}$ and $\log g$, the central star of PN~G325.3$-$02.9 falls 
outside the expected range for post-AGB stars, pointing to a very low mass object, with 
$M_{\rm CSPN}\lesssim 0.5M_{\odot}$. If this is the case, then the central star of 
PN~G325.3$-$02.9 could be the descendent of a former hot subdwarf star that avoided the 
AGB phase, {which implies that  the surrounding nebula cannot be} a bona~fide PN.
Due to the relatively low temperatures and high gravities derived from {our} 
spectra,  {the four CSPNe} are consistent with 
low masses and large post-AGB ages. 
In {particular,} the latter are well beyond the expected lifetimes of PNe, 
rising questions about {either}
 the actual status of the {PNe,} or the accuracy of the parameters derived 
from our spectroscopy. Concerning the first point, perhaps one can legitimately wonder if 
all the objects under study are bona~fide PNe: after all, all of them have received little
attention so far; perhaps the ionized gas is interstellar material and not the
star's ejection, and the star is not a post-AGB one at all,
or perhaps the central star has suffered more than one mass ejection, making
the object appear like a younger PN, masking its true age. {It is known,
for example, that 
the presence and evolution of a binary
at the center of a PN may affect the nebular morphology (\object[PN Fg 1]{Fg~1}, \citealt{bof12}),
and even produce double shells
(\object[A66 65]{A66~65}, \citealt{huc13}).}
About the second point, {we note that our spectra  are not 
of high quality, and that the  comparisons with the TheoSSA  models}
are based on the fitting of just \emph{one} line, assuming a certain chemical composition. 
For these {reasons, we can admit} that the values of $T_\mathrm{eff}$ and
$\log g$ we used {are} debatable. However, when we search  the literature
for estimated  $T_\mathrm{eff}$ and $\log g$ for other PNe,
{it becomes apparent that our parameters} are not so {out of place among them}. In Table~\ref{tab:literatura},
where we put together data for thirteen CSPNe from different sources, our objects
{fall in, or near,} the ranges of parameters listed.

\begin{deluxetable*}{lccccc} 
\tablecaption{Properties of CSPNe derived from
  interpolation/extrapolation of the $Z=0.01$ 
  stellar evolution models  presented by \citet{mm16}
\label{tab:CSPNe-derived}}
\tablewidth{0pt}
\tablehead{
  \colhead{PN~G} &
  \colhead{$\log L_{\rm CSPN}/L_\odot$\tablenotemark{a}} &
  \colhead{Age\tablenotemark{b}} &
  \colhead{Age$_{\rm Min}$\tablenotemark{c}} &
\colhead{Mass\tablenotemark{d}}  &
\colhead{$(M_{\rm min},M_{\rm max})$\tablenotemark{e}} \\
 & & (yr) & (yr) & ($M_\odot$) & ($M_\odot$)
} 
\startdata
019.7$-$10.7    & $1.77^{+0.32}_{-0.31}$  & 2.5$\times 10^{5}$ & 1.4$\times 10^{5}$  & 0.508      & $(\lesssim 0.5,0.538)$       \\
237.0$+$00.7    & $0.87^{+0.31}_{-0.28}$  & 6.9$\times 10^{5}$ & 3.1$\times 10^{5}$  & 0.543      & $(\lesssim 0.5,0.628)$       \\
276.2$-$06.6    & $1.43^{+0.31}_{-0.30}$  & 2.9$\times 10^{5}$ & 0.7$\times 10^{5}$  & 0.532      & $(\lesssim 0.5,0.596)$       \\
\enddata
\tablenotetext{a}{Interpolated CSPN luminosity.} 
\tablenotetext{b}{Interpolated post-AGB age.} 
\tablenotetext{c}{Minimum post-AGB age consistent with the models within the error ellipse.} 
\tablenotetext{d}{Interpolated mass.} 
\tablenotetext{e}{Interpolated mass range consistent  with the models within the error ellipse.}
 \end{deluxetable*}

 \begin{deluxetable*}{lccccl} 
\tablecaption{Parameters for planetary nebulae in the literature. 
This table includes the 30\% of known WDs
\label{tab:literatura}}
\tablewidth{0pt}
\tablehead{
\colhead{Usual name} & 
\colhead{PN~G} &
\colhead{Sp.~Type}  &
\colhead{$T_{\mathrm{eff}}$} &
\colhead{$\log g$} &
\colhead{Reference} \\
 & & & ($10^3$~K) & (cm~s$^{-2}$) &
} 
\startdata
\object[Sh 2-68]{Sh~2-68} & 030.6$+$06.2 &  hybrid & 84 & 7.24 &   \cite{2010ApJ...720..581G} \\
\object[NGC 7293]{NGC~7293} & 036.1$-$57.1 &  DAO    & 90 & 6.90 &   \citet{mkk1992}  \\
\object[NGC 6853]{NGC~6853} & 060.8$-$03.6 &  DAO    & 87 & 7.36 &   \cite{2010ApJ...720..581G}  \\
\object[NGC 6720]{NGC~6720} & 063.1$+$13.9 &  hgO(H) & 101 & 6.90 &  \citet{gde2013}  \\
\object[PN A66 61]{A66~61} & 077.6$+$14.7 &  DAO    & 88 & 7.10 &   \citet{napi1999}  \\
\object[PN G128.0$-$04.1]{Sh~2-188} & 128.0$-$04.1 &  DAO    & 87 & 7.41 &   \cite{2010ApJ...720..581G}  \\
\object[NGC 3587]{NGC~3587} & 148.4$+$57.0 &  DAO    & 94 & 6.90 &   \citet{gde2013}  \\
\object[PN HDW 3]{HDW~3} & 149.4$-$09.2 &  DAO    & 91 & 7.32 &  \cite{2010ApJ...720..581G}  \\
\object[PN PuWe 1]{PuWe~1} & 158.9$+$17.8 &  DAO    & 94 & 7.10 &  \citet{gde2013}  \\
\object[PN A66 7]{A66~7} & 215.5$-$30.8 &  DAO    & 99 & 7.00 &  \citet{gde2013}  \\
\object[EGB 6]{EGB~6} & 221.5$+$46.3 &  DAOZ   & 100 & 7.00 &  \citet{gpo1988}  \\
\object[PN A66 35]{A66~35} & 303.6$+$40.0 &  DAO    & 80 & 7.20 &   \citet{zie12}  \\
\enddata
\end{deluxetable*}

\newpage


\section{Summary and conclusions}
\label{sec:sum}
We have presented a method for identifying  white dwarfs as nuclei of planetary nebulae. 
The first results of this procedure have been very satisfactory. Using Gemini-GMOS
images and spectra, we
 have been able to find and characterize the ionizing sources of seven old nebulae, 
PN~G019.7$-$10.7, PN~G237.0$+$00.7, PN~G276.2$-$06.6, PN~G298.7$-$07.5,
PN~G302.1$+$00.3,
 PN~G314.5$-$01.0, and
 PN~G325.3$-$02.9.
Based on the analysis of our spectra and the modelling of the stellar atmospheres, 
we conclude that four of these stars are white dwarfs of  type DAO, that
another two are also likely WDs, and that cannot  discard  the last one as a WD as well.
{Another two possible CSPNe, those of PN~G328.5$+$06.0 and
PN~G344.9$+$03.0, have also been identifed based on photometric data.} 
Parameters for each CSPN,  including \emph{Gaia}-DR2 distances and colors, 
have been collected in Table~\ref{tab:resumen_de_parametros}.

These objects have colors $(u'-g')$ and $(g'-r')$ that are compatible with 
those of WDs, and they are  at the geometric center of PNe of large angular sizes 
and low surface brightness (i.e., old PNe). The spectra allowed us to classify
the  stars  as WDs.
The FWHM of absorption lines is compatible with the those of WD's spectra.
{\it Gaia} colors (allowing for reddening) indicate that they are blue objects.
The TheoSSA atmosphere models permitted us to derive
$T_{\mathrm{eff}}$ and $\log g$, obtaining  values that are typical of  
white dwarfs and not of hot subdwarfs.
We conclude, without doubt, that we have identified {at least} six white dwarfs---four
of them of the DAO subtype---as
 the ionizing sources of planetary nebulae.

\begin{deluxetable*}{lcccccc} 
\tablecaption{Summary of parameters for the CSPNe found in this work
\label{tab:resumen_de_parametros}}
\tablewidth{0pt}
\tablehead{
\colhead{PN~G} &
\colhead{$T_{\mathrm{eff}}$} &
\colhead{$\log g$} &
\colhead{Sp.\ Type} &
\colhead{\emph{G}} &
\colhead{$G_\mathrm{BP} - G_\mathrm{RP}$} &
\colhead{Distance} \\
 & ($10^3$~K) & (cm~s$^{-2}$) & & (mag) & (mag) & (kpc)
}
\startdata
019.7$-$10.7 &   83 &  7.00 & DAO  & 18.452 & $-0.481$ & 1.83 \\
237.0$+$00.7 & 65 & 7.50 & DAO & 18.128 & $-0.342 $ & 1.43 \\  
276.2$-$06.6 & 80 & 7.30  & DAO & 18.762 & $-0.317$ & 1.45 \\  
298.7$-$07.5 & \nodata  & \nodata  & WD? & 19.973 & $-0.330$ & 3.90       \\
302.1$+$00.3 & \nodata  & \nodata  & WD? & 19.407 & $+0.401$ & 2.60 \\
314.5$-$01.0 & \nodata     & \nodata & cont. & 18.528 & $+1.087$        &  2.00 \\ 
325.3$-$02.9 & 60 & 6.75 & DAO & 16.690 & $+0.424$ & \nodata \\
328.5$+$06.0 & \nodata & \nodata & \nodata &  18.698  & $+0.161$ & 1.98  \\
344.9$+$03.0 & \nodata & \nodata & \nodata &19.027  & $+0.328$ & 2.26 \\
\enddata
\end{deluxetable*}

We are currently applying this procedure to a number of
similar PNe  observed with the same telescope and instrument. Our goal is to extend the 
sample of known WDs as CSPNe in a 50\%, i.e.,
to identify and classify at least  fifteen new stars
of this kind.

We believe that it is important 
 to increase the known number of WDs that are, at the same time, CSPNe, as well as 
 to improve their spectral classification.
 This  will result 
in a refinement of the evolutionary models for the
progenitors of the PNe and, consequently, in a better
understanding of these fascinating objects.

\acknowledgments

Based on observations obtained at the Gemini Observatory and processed 
using the Gemini IRAF package. The Gemini Observatory  is operated by 
the Association of Universities for Research in Astronomy, Inc., under 
a cooperative agreement with the NSF on behalf of the Gemini partnership: 
the National Science Foundation (United States), the National Research 
Council (Canada), CONICYT (Chile), Ministerio de Ciencia, Tecnolog\'{i}a 
e Innovaci\'{o}n Productiva (Argentina), and Minist\'{e}rio da Ci\^{e}ncia, 
Tecnologia e Inova\c{c}\~{a}o (Brazil).
This research was made possible through the use of the AAVSO Photometric 
All-Sky Survey (APASS), funded by the Robert Martin Ayers Sciences Fund. 
W.~W.\ would like to thank  S.~O.~Kepler Oliveira and {R.~H.~M\'{e}ndez}
for their useful comments.
Part of this research was supported 
by grant SeCyT UNC project NRO PIP: 33820180100080CB.
This work has made use of data from the European Space Agency (ESA) mission
{\it Gaia} (\url{https://www.cosmos.esa.int/gaia}), processed by the {\it Gaia}
Data Processing and Analysis Consortium (DPAC,
\url{https://www.cosmos.esa.int/web/gaia/dpac/consortium}). Funding for the 
DPAChttps://es.overleaf.com/project/5c94e7484b4970168480de80
has been provided by national institutions, in particular the institutions
participating in the {\it Gaia} Multilateral Agreement.
The TMAW tool (\url{http://astro.uni-tuebingen.de/~TMAW}) used for this paper was 
constructed as part of the activities of the German Astrophysical Virtual Observatory.
{This research has made use of NASA's Astrophysics Data System and
the SIMBAD database, operated at CDS, Strasbourg, France. Finally, we would like to thank
the anonymous reviewer, whose comments and suggestions greatly helped to
 improve this paper.}

\vspace{5mm}
\facilities{GEMINI:South(GMOS)}

\software{Aladin \citep{bo00}, IRAF \citep{to93},  TMAW \citep{2018MNRAS.475.3896R}, 
\textsc{TopCat}  \citep{topcat}}


\newpage

\appendix

\section{FINDING CHARTS FOR THE CSPN}
\label{app:apendice}

\begin{figure*} 
\centering
\includegraphics[scale=3.]{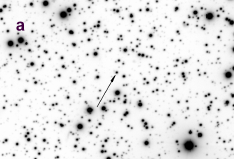}
\includegraphics[scale=3.]{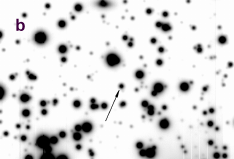}

\includegraphics[scale=3.]{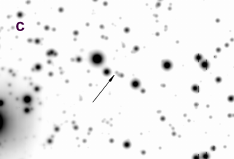}
\includegraphics[scale=3.]{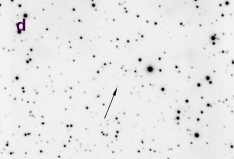}

\includegraphics[scale=3.]{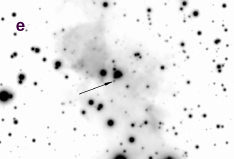}
\includegraphics[scale=3.]{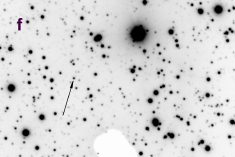}

\includegraphics[scale=3.]{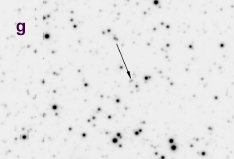}

\caption{Finding charts, adapted from our $r'$ images, 
 for the proposed central stars of the 
 planetary nebulae 
 PN~G019.7$-$10.7 (a), 
237.0$+$00.7 (b), 
276.2$-$06.6 (c), 
298.7$-$07.5 (d),  
302.1$+$00.3 (e), 
314.5$-$01.0  (f), and  
325.3$-$02.9  (g). 
The stars
 are marked by arrows and their coordinates are in 
 Table~\ref{tab:np}. 
 All charts cover areas of $3' \times 2'$. North is up and
 East is to the left.
 \label{fig:cartas}}
\end{figure*}

\begin{figure*} 
\centering
\includegraphics[scale=3.]{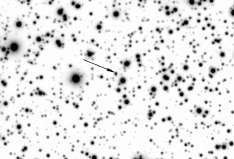}
\includegraphics[scale=3.]{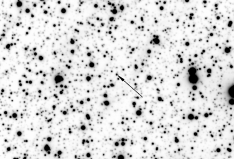}
\caption{The same as in Fig.~\ref{fig:cartas},
for PN~G328.5$+$06.0 (\emph{left}) and PN~G344.9$+$03.0 (\emph{right}).
 \label{fig:cartas2}}
\end{figure*}

\newpage

\bibliography{sample63}{}
\bibliographystyle{aasjournal}

\end{document}